\documentclass[nofootinbib,reprint,amsmath,amssymb,,showpacs,showkeys,aps]{revtex4-1}
\usepackage{graphicx}
\usepackage[utf8,latin1]{inputenc}
\usepackage{dcolumn}
\usepackage{setspace}
\usepackage{bm}
\usepackage{bigints}
\usepackage{hyperref}
\usepackage{amsmath}
\newcommand{\qm}[1]{``#1''}
\usepackage{hyperref}
\hypersetup{colorlinks, linkcolor={red},citecolor={blue},urlcolor={blue}}

\begin{document}

\preprint{APS/123-QED}

\title[General relativistic Poynting-Robertson to diagnose wormholes existence]{General relativistic Poynting-Robertson effect\\ to diagnose wormholes existence:\\ Static and spherically symmetric case}

\author{Vittorio De Falco$^{1}$}\email{vittorio.defalco@physics.cz}
\author{Emmanuele Battista$^{2,3}$}\email{ emmanuele.battista@kit.edu}
\author{Salvatore Capozziello$^{4,5}$}\email{capozzie@na.infn.it}
\author{Mariafelicia De Laurentis$^{4,5}$\vspace{0.5cm}}\email{mariafelicia.delaurentis@unina.it}

\affiliation{$^1$ Research Centre for Computational Physics and Data Processing, Faculty of Philosophy \& Science, Silesian University in Opava, Bezru\v{c}ovo n\'am.~13, CZ-746\,01 Opava, Czech Republic\\
$2$ Institute for Theoretical Physics, Karlsruhe Institute of Technology (KIT), 76128 Karlsruhe, Germany\\
$3$ Institute for Nuclear Physics, Karlsruhe Institute of Technology (KIT), Hermann-von-Helmholtz-Platz 1, 76344 Eggenstein-Leopoldshafen, Germany\\
$^4$ Universit\`{a} degli studi di Napoli \qm{Federico II}, Dipartimento di Fisica \qm{Ettore Pancini}, Complesso Universitario di Monte S. Angelo, Via Cintia Edificio 6, 80126 Napoli, Italy\\
$^5$ Istituto Nazionale di Fisica Nucleare, Sezione di Napoli, Complesso Universitario di Monte S. Angelo, Via Cintia Edificio 6, 80126 Napoli, Italy}

\date{\today}

\begin{abstract}
We derive the equations of motion of a test particle in the equatorial plane around a static and spherically symmetric wormhole influenced by a radiation field including the general relativistic Poynting-Robertson effect. From the analysis of this dynamical system, we develop a diagnostic to distinguish a black hole from a wormhole, which can be timely supported by several and different observational data. This procedure is based on the possibility of having some wormhole metrics, which smoothly connect to the Schwarzschild metric in a small transition surface layer very close to the black hole event horizon. To detect such a metric-change, we analyse the emission proprieties from the critical hypersurface (stable region where radiation and gravitational fields balance) together with those from an accretion disk in the Schwarzschild spacetime toward a distant observer. Indeed, if the observational data are well fitted within such model, it immediately implies the existence of a black hole; while in case of strong departures from such description it means that a wormhole could be present. Finally, we discuss our results and draw the conclusions.
\end{abstract}

\maketitle

\section{Introduction}
\label{sec:intro}
Black holes (BHs) constitute one of the main implications of General Relativity (GR) and of any metric theory of gravity. The recent observational evidences about their existence and nature through the detection of gravitational waves (GWs) from the Laser Interferometer Gravitational-Wave Observatory (LIGO) \cite{Abott2016a,Abott2016b,Abott2017} and the first image of the BH located at the center of the Galaxy M87 from the Event Horizon Telescope (EHT) collaboration \cite{EHC20191,EHC20192,EHC20193,EHC20194,EHC20195,EHC20196} shed new light on this intriguing gravitational massive sources. Such astrophysical objects are well known to be characterized by an event horizon, which is a one way-membrane separating the smoothly-behaving exterior spacetime from the casually disconnected inner regions hiding essential singularities at their center.  

Actually, there are several techniques to gather information about such objects: the oldest practice consists in observing the interaction between BH and the surrounding matter through the emission of electromagnetic signals in the X-ray energy band \cite{Done2002}; tracking the motion of stellar objects around a supermassive BH (SBH), as it has been doing for long time with Sgr A* \cite{Gillessen2017}; the detection of GWs signals through instruments of higher and higher sensibilities \cite{Gourgoulhon2019,Abuter2020}; the ability of imaging the matter dynamics in the vicinity of a BH \cite{EHC20191,EHC20192,EHC20193,EHC20194,EHC20195,EHC20196,Kim2020}. This period, defined the \emph{multi-messenger era} for the wealth of complementary observational data, offers the possibility to have finally more insight into the BHs' description within or outside the GR theory. 

There is a huge variety of theoretical compact objects' candidates, which perfectly mimic all observational properties of a BH with arbitrary accuracy \cite{Cardoso2019}. In this huge class of BH mimickers, an appealing position is occupied by the wormholes (WHs) \cite{Damour2007,Bambi2012,Bambi2013}. They have the peculiar proprieties to be horizonless, and endowed with a traversable bridge connecting two different universes \cite{Visser1995}. The traversable condition in classical GR is linked to the existence of exotic matter, having negative energy and going against the classical laws of physics. A common way to explain such issue is based on quantum mechanics \cite{Hochberg1997,Bronnikov2013,Digrezia2017,Garattini2019} or, if we frame the WH models in alternative theories of gravity, on topological arguments \cite{Lobo2009,Harko2013}.

In the literature, several have been the attempts and new techniques proposed to detect WHs and establish their observational signatures. In order to give a precise idea of such research field, it is worth citing the following works: Cardoso and Pani \cite{Cardoso2016} noted that WHs with a light ring admit similar BH ringdown stage, and their quasinormal-mode spectrum, which is completely different from that of a BH, can eventually show up only at late times; instead Konoplya and Zhidenko \cite{Konoplya2016} showed later that particular classes of WHs can actually ring similarly or differently to BHs at all times; Paul and collaborators \cite{Paul2019} produced numerical images of a thin accretion disk around both a BH and WH, determining distinctive features when a WH has accretion disks on both sides of its throat, and qualitatively similar or dramatic differences when the disk is on the same side of the observer (see figures in the paper, for details); Dai and collaborators \cite{Dai2019} showed that the gravitational flux propagates from one universe to the other one perturbing the motion of the objects, detectable with an acceleration precision of $10^{-6}\ {\rm m/s^2}$; Banerjee and collaborators \cite{Banerjee2019} calculated that BHs and WHs close to the event horizon/throat have distinctive tidal effects (a few times higher for WHs) arising from their different geometries; Hashimoto and Dalui \cite{Hashimoto2017,Dalui2019} found that the motion of massive and massless test particles exhibit chaotic behaviors near the event horizon due to its surface gravity, permitting to probe whether there exists a horizon. 

In this work, we consider static and spherically symmetric WHs in pure GR, mimickers of BHs' proprieties. We propose an original procedure to diagnose a WH from a BH by employing the general relativistic Poynting-Robertson (PR) effect, which can be supported by the recent massive amount of observational data.

In high-energy astrophysics, the motion of relatively small-sized test particles (like accretion disk elements, meteors, comets, planets, dusts) around massive compact objects (like SBHs or stellar BHs) is influenced not only by the gravitational field, but also by the electromagnetic radiation from an emitting source (like accretion disk, or hot corona around a BH). Beside such forces, there is also the presence of a radiation drag force, termed PR effect, arising when the matter absorbs and reemits the radiation, generating thus a thrust force opposite to the matter orbital motion \cite{Ballantyne2004,Ballantyne2005,Worpel2013,Ji2014,Keek2014,Worpel2015}. Such effect configures as a dissipative force, which removes energy and angular momentum from the effected body \cite{Poynting1903,Robertson1937,Bini2009,Bini2011}. Recent works on such topic are: extension from the two-dimensional (2D) formulation to the three-dimensional (3D) space in Kerr metric \cite{Defalco20183d,Bakala2019}; continuous emission of radiation from a finite source in Schwarzschild spacetime \cite{Wielgus2019}; treatment under a Lagrangian formulation, where the Rayleigh potential (describing the radiation dissipative force) has been analytically determined for the first time in the GR literature \cite{DeFalco2018,Defalco2019,DeFalco2019VE}; proof within the Lyapunov theory that the critical hypersurfaces (regions where there is a balance between radiation and gravitational forces) are stable configurations \cite{Defalco2019ST}.

The paper is structured as follows: in Sec. \ref{sec:PReffect} we first derive the equations of motion of a test particle around a static and spherically symmetric WH affected by the general relativistic PR effect, discussing the proprieties and implications of such dynamics with respect to the Schwarzschild case; in Sec. \ref{sec:diagnostic} we present our proposal to disentangle a WH from a BH by analysing the electromagnetic emission proprieties from the critical hypersurfaces in Schwarzschild and WH spacetime under the PR effect; in Sec. \ref{sec:end} we discuss about the obtained results and finally give our conclusions.  

\section{General relativistic Poynting-Robertson effect around a static and spherically symmetric wormhole}
\label{sec:PReffect}
In this section, we recall the proprieties of a Morris-Thorne WH metric (see Sec. \ref{sec:MTmetric}), which will be the geometrical background on which a test particle will move under the general relativistic PR effect. After having derived its equations of motion (see Sec. \ref{sec:GRPReffect}), we analyse the existence of the critical hypersurfaces (see Sec. \ref{sec:CH}), one of the most important implications of the general relativistic PR effect, which we will use in the next sections. 

\subsection{The Morris--Thorne Wormhole}
\label{sec:MTmetric}
We consider a static and spherically symmetric WH, whose spacetime is described by the Morris--Throne metric \cite{Morris1988}. We use the signature $(-,+,+,+)$ for the metric, and geometrical units for gravitational constant $G$, and speed of light $c$ ($c = G = 1$). The metric line element, $ds^2=g_{\alpha\beta}dx^\alpha dx^\beta$, expressed in spherical coordinates and in the equatorial plane $\theta=\pi/2$, reads as 
\begin{equation} \label{eq:MTmetric}
ds^2=-e^{2\Phi(r)}dt^2+\frac{dr^2}{1-b(r)/r}+r^2d\varphi^2,
\end{equation}
which is parametrized by $\Phi(r)$ and $b(r)$, better known as the \emph{shape} and \emph{redshift functions}, respectively. 
\begin{figure*}
\centering
\includegraphics[scale=0.49]{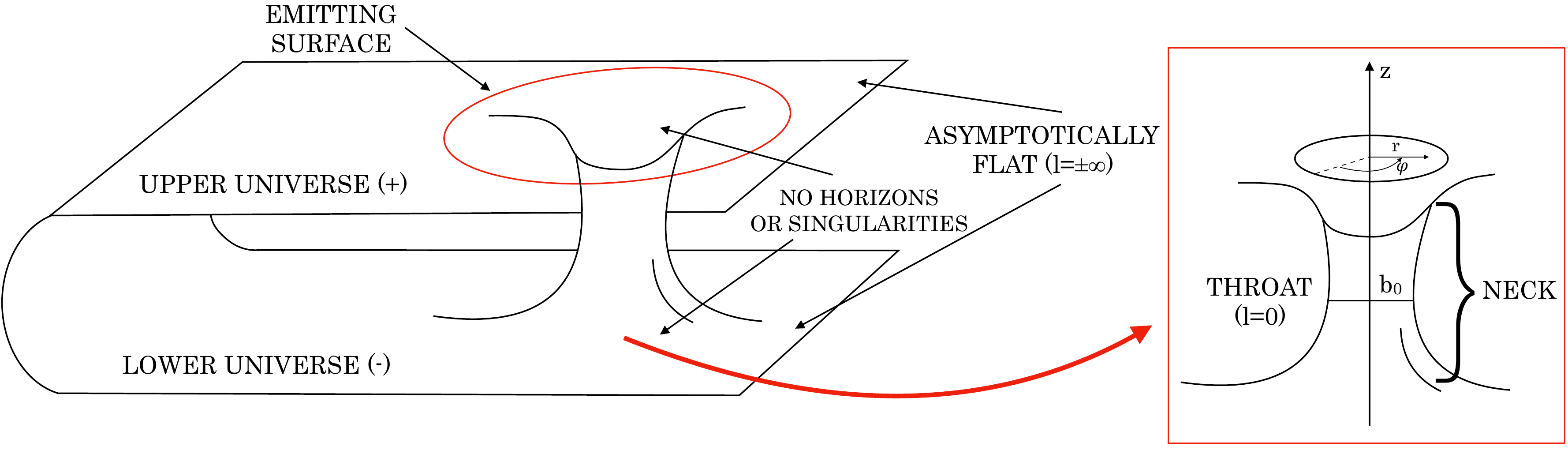}
\caption{Sketch of the Morris--Throne WH geometry including the presence of an accretion disk in the upper universe.}
\label{fig:Fig1}
\end{figure*}

To simplify the calculations and have a direct physical interpretations of the results we will find, we employ as orthonormal basis of vectors the proper reference frame adapted to the \emph{static observers} (SOs), given by \cite{Morris1988}
\begin{equation} \label{eq:SOframe}
\begin{aligned}
&\boldsymbol{e_{\hat t}}\equiv\boldsymbol{n}= \frac{\boldsymbol{\partial_t}}{N},\quad
\boldsymbol{e_{\hat r}}=\frac{\boldsymbol{\partial_r}}{\sqrt{g_{rr}}},\quad
\boldsymbol{e_{\hat \varphi}}=\frac{\boldsymbol{\partial_\varphi}}{\sqrt{g_{\varphi \varphi }}}.
\end{aligned}
\end{equation}
where $N\equiv(-g^{tt})^{-1/2}=e^{\Phi(r)}$ is the time lapse function \cite{Bini2009,Bini2011}. We will denote throughout the paper vector and tensor indices (e.g., $v_\alpha$; $T_{\alpha\beta}$) evaluated in the SO frame by a hat (e.g., $v_{\hat \alpha}$; $T_{\hat{\alpha}\hat{\beta}}$), instead scalar quantities (e.g., $f$) measured in SO frame will be followed by $n$ (e.g., $f(n)$).

The geometrical proprieties of a WH entail some constraints on the $b(r),\Phi(r)$ functions, which are \cite{Morris1988}:
\begin{itemize}
\item the presence of a spatial embedded 2D surface, connecting two spacetimes (defined as \emph{WH neck}, see Fig. \ref{fig:Fig1}). It is described in cylindrical coordinates by the following differential equation
\begin{equation} \label{eq:whshape}
\frac{dz(r)}{dr}=\pm\left[\frac{r}{b(r)}-1\right]^{-1/2},
\end{equation}
where the sign "+" stays for the upper universe, and the sign "-" is for the lower universe. The surface $z=z(r)$ gives the WH neck shape; 
\item \emph{there are no horizons or singularities}, entailing therefore that $\Phi$ is everywhere finite;
\item the \emph{WH throat} is defined as the minimum radius such that $r_{\rm min}=b_0$ and $b(r_{\rm min})=b_0$. Such definition substituted in Eq. (\ref{eq:whshape}) gives a divergence, being formally not in agreement with the previous point. However, by exploiting the proper radial distance $l$ as measured by SOs \cite{Morris1988}
\begin{equation}\label{eq:ldist2}
\frac{dl}{dr}=\pm\left[1-\frac{b(r)}{r}\right]^{-1/2},
\end{equation}
or in a clearer form as
\begin{equation}\label{eq:ldist2}
l(r)=\pm\bigintss_{b_0}^r\frac{dr}{\left[1-\frac{b(r)}{r}\right]^{1/2}},
\end{equation}
we immediately see how the divergence disappear. Indeed, imposing that such distance is finite through all the spacetime, we require that $1-b(r)/r\ge0$ throughout the spacetime;
\item both connected universes are \emph{asymptotically flat} far from the throat in both radial directions, namely $b(r)/r\to0$ and $\Phi\to0$ for $l\to\pm\infty$;
\item another fundamental point is to characterize the material, which generates the WH spacetime curvature. We start by defining its stress-energy tensor, as measured in the SO frame, given by
\begin{equation} \label{eq:setm}
 T_{\hat{t}\hat{t}}^{\rm (m)}=\rho(r),\quad  T_{\hat{r}\hat{r}}^{\rm (m)}=-\tau(r),\quad  T_{\hat{\varphi}\hat{\varphi}}^{\rm (m)}=p(r),
\end{equation} 
where $\rho$ is the mass-energy density, $\tau$ is the radial tension, and $p$ is the lateral pressure. Now, solving the Einstein field equations and defining $(\cdot)'\equiv d(\cdot)/dr=(1-b(r)/r)^{-1/2}d(\cdot)/dl$, we obtain
\begin{eqnarray}
&&\rho(r)=\frac{b'(r)}{8\pi r^2},\label{eq:rho}\\
&&\tau(r)=\frac{b(r)/r-2[r-b(r)]\Phi'(r)}{8\pi r^2},\label{eq:tau}\\
&&p(r)=\frac{r}{2}\left[(\rho(r)-\tau(r))\Phi'(r)-\tau'(r)\right]-\tau(r).\label{eq:p}
\end{eqnarray}
These equations are extremely important, because they closely relate the matter to the metric functions and viceversa. In particular, Eq. (\ref{eq:rho}) can be easily integrated, giving
\begin{equation}
b(r)=b(r_0)+\int_{b_0}^r8\pi x^2\rho(x)dx=2m(r),
\end{equation}
where we have defined 
\begin{equation}
m(r)=\frac{b_0}{2}+\int_{b_0}^r4\pi x^2\rho(x)dx,
\end{equation}
which is the effective mass contained in the sphere of radius $r$. In this equation, it is more understandable the role of the function $b(r)$, which is linked to the distribution of masses inside the WH. In particular at spatial infinity, we have \cite{Visser1995}
\begin{equation} \label{eq:totmass}
\lim_{r\to\infty}m(r)=\frac{b_0}{2}+\int_{b_0}^\infty4\pi x^2\rho(x)dx=M,
\end{equation}
where $M$ is the total mass of the system. 

The traversable propriety is expressed by the \emph{flaring out condition}, which entails $\tau(r)>\rho(r)$ at the throat or  that the dimensionless function
\begin{equation} \label{eq:csi}
\xi(r)=\frac{\tau(r)-\rho(r)}{|\rho(r)|}
\end{equation}
be non-negative at $r=b_0$, or also that $d^2r/dz^2>0$ at $r=b_0$, giving a particular constraint on the WH shape. All these implications physically translates in having a \emph{negative mass-energy density} inside the throat. This leads to a delicate issue regarding the existence of \emph{exotic matter}, which, albeit several proposals, is still matter of debate and seems to be forbidden by classical laws of physics, but accepted through a quantum field theory argument. However, as discussed in \cite{Hochberg19981,Hochberg19982,caplobo1,caplobo2}, it is possible to bypass this difficulty by considering modified theories of gravity where energy conditions are not violated because the additional (geometric) degrees of freedom behave as a fluid, whose energy density can be eventually negatively defined. In other words, standard  fluid matter retains its own
properties but violations are prevented by the improved field equations (see e.g., Refs. \cite{Visser1989,Barcelo1999,Bohmer2012,Capozziello2012,Bahamonde2016,Capozziello2020}, for further discussions and approaches). The debate is that exotic matter or modified gravity should be taken into account to obtain realistic WHs.

\end{itemize}
In this work we are interested only in the \emph{basic WH criteria}, caring only about the geometrical structure, without considering the \emph{usability criteria}, which are defined to tune the WH for human interstellar travels (e.g., traversability of the WH neck in a relatively short time period, comfortable radial tidal force) \cite{Morris1988}. 

\subsection{The Poynting-Robertson effect in General Relativity}
\label{sec:GRPReffect}
In this section, we aim at deriving the equations of motion of a test particle influenced by the gravitational force from the WH, the radiation pressure together with the general relativistic PR effect from a radiation source outside the WH throat. We adopt the following strategy: we first write the equations of motion in the SO frame, see Eq. (\ref{eq:SOframe}), and then transform them in the frame of the static observer at infinity, see Eq. (\ref{eq:MTmetric}). To this end, we make use of the \emph{observer splitting formalism}, which is able to coherently disentangle gravitational from fictitious forces arising from the relative motion of two non-inertial observers \cite{Jantzen1992,Bini1997a,Bini1997b,DeFalco2018}.

We calculate the SO kinematical quantities, which are acceleration $\boldsymbol{a}(n)=\nabla_{\boldsymbol{n}} \boldsymbol{n}$, and the relative Lie curvature vector $\boldsymbol{k_{(\rm Lie)}}(n)$, whose explicit expressions are \cite{Bini2009,Defalco20183d}
\begin{eqnarray}
a(n)^{\hat r}&=&\Phi'(r)\sqrt{1-b(r)/r},\\
k_{\rm (Lie)}(n)^{\hat r}&=&-\frac{\sqrt{1-b(r)/r}}{r}.
\end{eqnarray}

\subsubsection{Radiation field}
\label{sec:radfield}
We model the radiation field as a coherent flux of photons traveling along null geodesics on the Morris-Thorne metric. The photons depart from a radiation source around the WH (see Fig. \ref{fig:Fig1}), but located outside the neck, and only one single photon reaches the test particle at its position at each instant of time. In this case, it is important to underline that the radiation stress-energy tensor $T^{\mu\nu}$ is superimposed on the background geometry, without modifying it. Such tensor is different from the one occurring in Eq. (\ref{eq:setm}), where we have used the superscript \qm{$(m)$}, and reads as \cite{Bini2009,Bini2011}
\begin{equation}\label{eq:SET}
T^{\mu\nu}=\mathcal{I}^2 k^\mu k^\nu\,,\qquad k^\mu k_\mu=0,\qquad k^\mu \nabla_\mu k^\nu=0,
\end{equation}
where $\mathcal{I}$ is a parameter linked to the radiation field intensity, $\boldsymbol{k}$ is the photon four-momentum field, and the last two equations express the null geodesic condition. In such spacetime, we have that the energy $E=-k_t$, and the angular momentum with respect to the polar axis (or whatever other axis) $L_z=k_\varphi$, are conserved quantities along the photon trajectory. Splitting $\boldsymbol{k}$ with respect to the SO frame, we obtain \cite{Bini2009,Bini2011}
\begin{eqnarray}
&&\boldsymbol{k}=E(n)[\boldsymbol{n}+\boldsymbol{\hat{\nu}}(k,n)], \label{photon1}\\
&&\boldsymbol{\hat{\nu}}(k,n)=\sin\beta\ \boldsymbol{e_{\hat r}}+\cos\beta\ \boldsymbol{e_{\hat\varphi}}, \label{photon2}
\end{eqnarray}
where 
\begin{equation}
E(n)=\frac{E}{e^{\Phi(r)}},
\end{equation}
is the photon energy measured in the SO frame, $\boldsymbol{\hat{\nu}}(k,n)$ is the photon spatial unit relative velocity with respect to the SO frame, $\beta$ is the angle measured in the SO frame in the azimuthal direction. The radiation field is governed by the impact parameter $\lambda=L_z/E$, associated with the emission angle $\beta$. The radiation field photons are emitted from a spherical rigid surface having a radius $R_\star$ centered at the origin of the spherical coordinates, and rotating rigidly with angular velocity $\Omega_{\mathrm{\star}}$. 
The photon impact parameter $b$ and the related photon angle $\beta$ have the following expressions \cite{Bini2011,Bakala2019}
\begin{equation} \label{MT_impact_parameter}
\lambda=\Omega_{\star}\left[\frac{\mathrm{g_{\varphi\varphi}}}{-\mathrm{g_{tt}}}\right]_{r=R_\star},\quad \cos\beta=\frac{e^{\Phi(r)}}{r}\lambda, 
\end{equation}
where we indicate with the label $r=R_\star$ that the metric components $g_{\varphi\varphi},g_{tt}$ must be evaluated in $R_\star$.

From the conservation of the stress-energy tensor, namely $\nabla_\mu T^{\mu\nu}=0$, we are able to determine the parameter $\mathcal{I}$, which has the following expression \cite{Bini2009,Bini2011}
\begin{equation}\label{INT_PAR}
\mathcal{I}^2=\frac{\mathcal{I}_0^2}{\sqrt{r^2-e^{2\Phi(r)}\lambda^2}},
\end{equation}
where $\mathcal{I}_0$ is $\mathcal{I}$ evaluated at the emitting surface.

\subsubsection{Radiation force and test particle acceleration}
\label{sec:radforce}
A test particle moves with a timelike four-velocity $\boldsymbol{U}$ and a spatial velocity $\boldsymbol{\nu}(U,n)$ with respect to the SO frames, which both read as \cite{Bakala2019}
\begin{eqnarray} 
&&\boldsymbol{U}=\gamma(U,n)[\boldsymbol{n}+\boldsymbol{\nu}(U,n)], \label{testp}\\
&&\boldsymbol{\nu}=\nu(\sin\alpha\boldsymbol{e_{\hat r}}+\cos\alpha \boldsymbol{e_{\hat\varphi}}),
\end{eqnarray}
where $\gamma(U,n)\equiv\gamma=1/\sqrt{1-||\boldsymbol{\nu}(U,n)||^2}$ is the Lorentz factor, $\nu=||\boldsymbol{\nu}(U,n)||$ is the magnitude of the test particle spatial velocity, and $\alpha$ is the azimuthal angle of the vector $\boldsymbol{\nu}(U,n)$ measured clockwise from the positive $\hat\varphi$ direction in the $\hat{r}-\hat{\varphi}$ tangent plane in the SO frame. 

The test particle acceleration $\boldsymbol{a}(U)$, can be calculated by using the relativity of observer splitting formalism. They can be easily derived in the SO frame by employing the proprieties of spherical symmetry shared with the Schwarzschild equations \cite{Bini2011}, i.e.,
\begin{eqnarray}
a(U)^{\hat t}&=&\gamma^2\nu\sin\alpha\ a(n)^{\hat r}+\gamma^3\nu\frac{d\nu}{d\tau},\\
a(U)^{\hat r}&=&\gamma^2[a(n)^{\hat r}+k_{\rm (Lie)}(n)^{\hat r}\nu^2\cos^2\alpha]\notag\\
&&+\gamma\left(\gamma^2\sin\alpha\frac{d\nu}{d\tau}+\nu\cos\alpha\frac{d\alpha}{d\tau}\right),\\
a(U)^{\hat \varphi}&=&-\gamma^2\nu^2\sin\alpha\cos\alpha k_{\rm (Lie)}(n)^{\hat r}\notag\\
&&+\gamma\left(\gamma^2\cos\alpha\frac{d\nu}{d\tau}-\nu\sin\alpha\frac{d\alpha}{d\tau}\right).
\end{eqnarray}
We assume that the radiation-test particle interaction occurs through Thomson scattering, characterized by a constant momentum-transfer cross section $\sigma$, independent of direction and frequency of the radiation field. We can split the photon four-momentum (\ref{photon1}) in terms of the velocity $\boldsymbol{U}$ as \cite{Bini2009,Bini2011,Defalco20183d,Bakala2019}
\begin{equation}
\boldsymbol{k}=E(U)[\boldsymbol{U}+\boldsymbol{\hat{\mathcal{V}}}(k,U)],
\end{equation}
where 
\begin{equation}
E(U)=\gamma E(n)[1-\nu\cos(\alpha-\beta)],
\end{equation}
is the photon energy measured by the test particle. The radiation force $\boldsymbol{{\mathcal F}_{\rm (rad)}}(U)$ can be written as \cite{Bini2009,Bini2011,Defalco20183d,Bakala2019}
\begin{equation} \label{radforce}
{\mathcal F}_{\rm (rad)}(U)^{\hat \alpha}=\tilde{\sigma} \, [\mathcal{I} E(U)]^2\, \hat{\mathcal V}(k,U)^{\hat \alpha},
\end{equation}
where $\tilde{\sigma}=\sigma/m$ and $m$ is the test particle mass. The term $\tilde{\sigma}[\mathcal{I} E(U)]^2$ has the following expression \cite{Bakala2019} 
\begin{equation} \label{eq: sigma_tilde}
\tilde{\sigma}[\mathcal{I} E(U)]^2=\frac{ A\,\gamma^2 [1-\nu\cos(\alpha-\beta)]^2}{e^{2\Phi(r)}\sqrt{r^2-e^{2\Phi(r)}\lambda^2}},
\end{equation}
where $A=\tilde{\sigma}[\mathcal{I}_0 E]^2$ being the luminosity parameter, which can be equivalently written as $A/M=L/L_{\rm EDD}\in[0,1]$ with $M$ is the mass defined in Eq. (\ref{eq:totmass}), $L$ the emitted luminosity at infinity, and $L_{\rm EDD}$ the Eddington luminosity. The terms $\hat{\mathcal V}(k,U)^{\hat \alpha}$ are the radiation field components, whose expressions are \cite{Bini2011,Bakala2019}
\begin{eqnarray}\label{rad}
&&\hat{\mathcal{V}}^{\hat r}=\frac{\sin\beta}{\gamma [1-\nu\cos(\alpha-\beta)]}-\gamma\nu\sin\alpha, \\ 
&&\hat{\mathcal{V}}^{\hat\varphi}=\frac{\cos\beta}{\gamma [1-\nu\cos(\alpha-\beta)]}-\gamma\nu\cos\alpha,\\
&&\hat{\mathcal{V}}^{\hat t}=\gamma\nu\left[\frac{\cos(\alpha-\beta)-\nu}{1-\nu\cos(\alpha-\beta)}\right].
\end{eqnarray}

\subsubsection{Equations of motion}
\label{sec:eom}
Collecting all the information derived in the previous sections, we are able to derive the equations of motion of a test particle moving in the equatorial plane around a WH and influenced by the radiation force (\ref{radforce}). Imposing that $\boldsymbol{a}(U)=\boldsymbol{{\mathcal F}_{\rm (rad)}}(U)$, we obtain \cite{Bini2009,Bini2011,Defalco20183d,Bakala2019}
\begin{eqnarray}
\frac{d\nu}{d\tau}&=& -\frac{\sin\alpha}{\gamma}a(n)^{\hat r}\label{EoM1}\\
&&+\frac{ A [1-\nu\cos(\alpha-\beta)][\cos(\alpha-\beta)-\nu]}{e^{2\Phi(r)}\sqrt{r^2-e^{2\Phi(r)}\lambda^2}},\nonumber\\
\frac{d\alpha}{d\tau}&=&-\frac{\gamma\cos\alpha}{\nu}\left[a(n)^{\hat r}+k_{\rm (Lie)}(n)^{\hat r}\,\nu^2\right]\label{EoM2}\\
&&+\frac{ A [1-\nu\cos(\alpha-\beta)]\sin(\alpha-\beta)}{e^{2\Phi(r)}\sqrt{r^2-e^{2\Phi(r)}\lambda^2}\ \nu\cos\alpha},\nonumber\\
U^{\hat r}&\equiv&\frac{dr}{d\tau}=\frac{\gamma\nu\sin\alpha}{\sqrt{g_{rr}}}, \label{EoM3}\\
U^{\hat \varphi}&\equiv&\frac{d\varphi}{d\tau}=\frac{\gamma\nu\cos\alpha}{\sqrt{g_{\varphi\varphi}}},\label{EoM4}\\
U^{\hat t}&\equiv&\frac{dt}{d\tau}=\frac{\gamma}{N},\label{time}
\end{eqnarray}
where $\tau$ is the affine parameter (proper time) along the test particle trajectory. 

\subsection{Critical hypersurfaces}
\label{sec:CH}
The dynamical system given by Eqs. (\ref{EoM1}) -- (\ref{EoM4}) may admit the existence of a critical hypersurface, a region where there is a balance between the radiation and gravitational forces. We already know that in the equatorial plane of the Schwarzschild metric \cite{Bini2009,Bini2011}, they behave as \emph{stable attractors}, namely particular configurations where the test particle moves stably on it for all future times. 

\begin{figure}[ht!]
\centering
\includegraphics[scale=0.315]{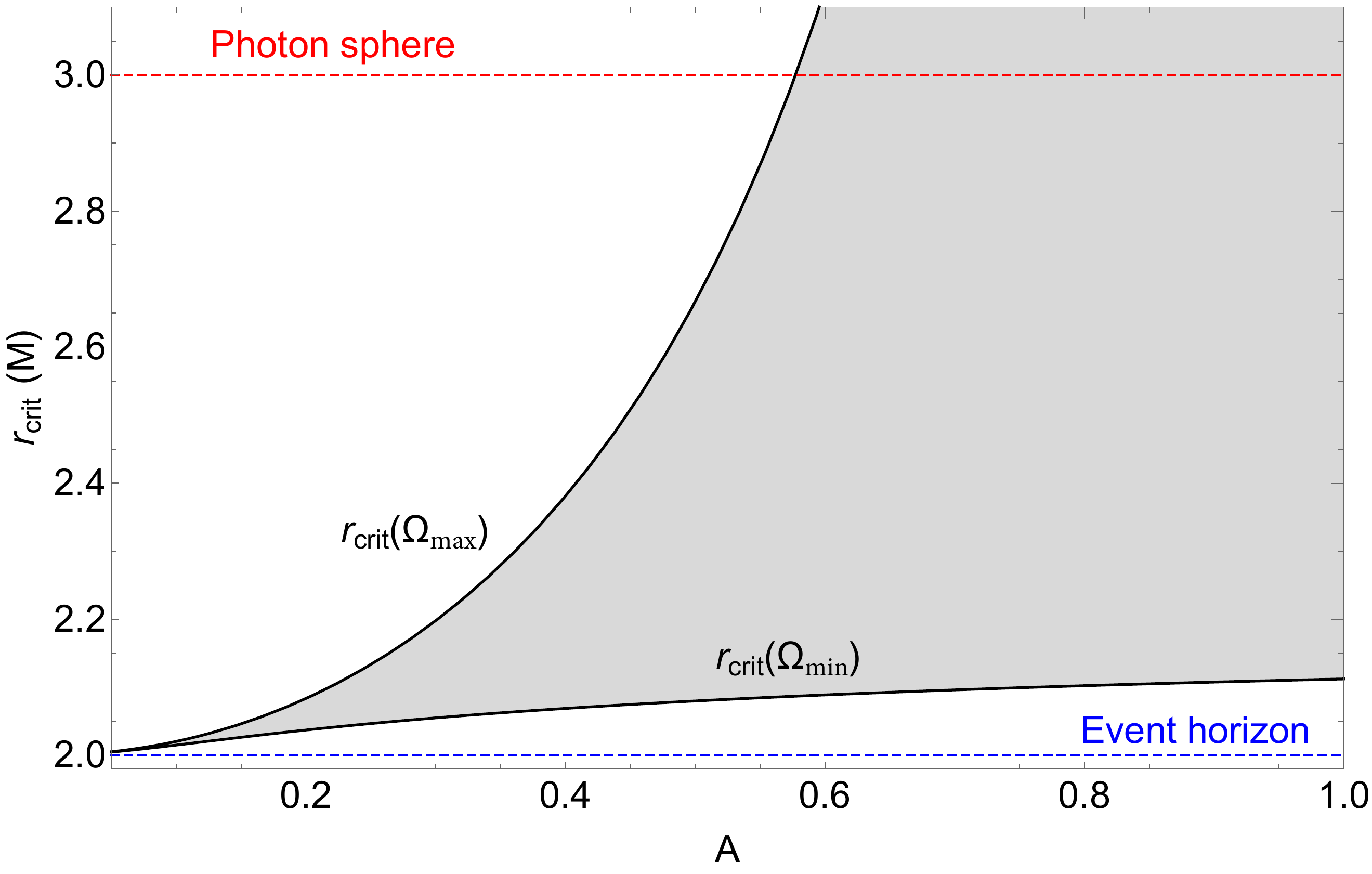}
\caption{Critical radius $r_{\rm crit}$ in terms of the luminosity parameter $A$. The grey region, delimited by the critical radius profiles evaluated at the minimum ($\Omega_{\rm min}=0$) and maximum ($\Omega_{\rm max}=[\sqrt{-g_{tt}/g_{\varphi\varphi}}]_{r=R_\star}$) angular frequencies, defines the region where all the critical radii vary. The dashed red line represents the photon sphere, while the dashed blue line the event horizon.}
\label{fig:Fig4}
\end{figure}

Imposing that on the critical hypersurface the test particle must move on purely circular orbits (i.e., $\alpha=0,\pi$) and with constant velocity (i.e., $\nu=\rm{const}$), we have that $d\nu/d\tau=0$, and $d\alpha/d\tau=0$, or equivalently that \cite{Bini2009,Bini2011}
\begin{eqnarray}
&&\frac{A [1-\nu\cos(\alpha-\beta)][\cos(\alpha-\beta)-\nu]}{e^{2\Phi(r)}\sqrt{r^2-e^{2\Phi(r)}\lambda^2}}=0,\label{eq:CH1}\\
&&a(n)^{\hat r}+k_{\rm (Lie)}(n)^{\hat r}\,\nu^2\notag\\
&&+\frac{A [1-\nu\cos\beta]\sin\beta}{\gamma e^{2\Phi(r)}\sqrt{r^2-e^{2\Phi(r)}\lambda^2}}=0. \label{eq:CH2}
\end{eqnarray}
From Eq. (\ref{eq:CH1}), we obtain that the velocity of the test particle on the critical hypersurface must be equal to the azimuthal photon velocity 
\begin{equation}
\nu=\cos\beta.
\end{equation}
Substituting such result in Eq. (\ref{eq:CH2}), we derive an implicit equation for determining the radius $r_{\rm crit}$ of the critical hypersurface, which is given by
\begin{equation} \label{eq:CH3}
a(n)^{\hat r}+k_{\rm (Lie)}(n)^{\hat r}\cos^2\beta+\frac{A \sin^3\beta}{r\ e^{2\Phi(r)}}=0.
\end{equation}
We already know the proprieties of Eq. (\ref{eq:CH3}) in the Schwarzschild metric. For photons emitted radially (i.e., $\lambda=0$), the test particle reaches the critical hypersurface and ends its motion on a point, since there is a perfect balance between radiation and gravitational forces \cite{Bini2009,Defalco20183d}; instead in the case where the photons are emitted in an arbitrary direction (i.e., $\lambda\neq0$), the test particle when reaches the critical hypersurface it starts to move on it with constant velocity given by Eq. (\ref{eq:CH1}) \cite{Bini2011,Bakala2019}. In addition, we know that the critical radius solution of Eq. (\ref{eq:CH3}) is unique\footnote{In the Schwarzschild case, Eq. (\ref{eq:CH3}) can admit three different solutions. One solution is located very far from the BH and another one is close to the event horizon, so they are unphysical. Therefore, there exists only one physical solution \cite{Bini2011,Bakala2019}.}, and it continuously depends on the luminosity parameter $A$. Therefore, it is always possible to find a critical radius very close to the Schwarzschild event horizon for a particular luminosity parameter $A$ (crucial propriety which we will be exploited in Sec. \ref{sec:diagnostic}), as one can immediately see in Fig. \ref{fig:Fig4}. 

\subsubsection{WH critical hypersurfaces: general remarks}
\label{sec:generalCH}
In our case, Eq. (\ref{eq:CH3}) is generic, being determined once we have an explicit expression of the metric functions $\Phi(r), b(r)$. Therefore, depending on the considered  WH, we can have different dynamical systems and it may happen that Eq. (\ref{eq:CH3}) admits more than one (physical) solution, or no solutions (in the worst case), or admits solutions only in some regions of the spacetime. Let us consider a particular case, where the temporal metric component is a constant function, i.e., $\Phi(r)=\Phi_0\equiv\rm{const}$, while the shape function $b(r)$ remains still unspecified. In such particular case, Eq. (\ref{eq:CH3}) becomes
\begin{equation} \label{eq:CH4}
\lambda^2e^{4\Phi_0}r\sqrt{1-\frac{b(r)}{r}}=A(r^2-e^{2\Phi_0}\lambda^2)^{3/2}.
\end{equation}
In this case, when the photons are radially emitted (i.e., $\lambda=0$), Eq. (\ref{eq:CH4}) implies $r=0$, which is not a physical solution neither for a BH nor for a WH, independently of the explicit functional form of the shape function $b(r)$. Therefore, for $\lambda=0$ it is never possible to have critical hypersurfaces. Instead, for $\lambda\neq0$, Eq. (\ref{eq:CH4}) becomes an algebraic equation in $r$ of sixth order,
\begin{equation} \label{eq:CH5}
\begin{aligned}
&A^2r^6-3e^{2\Phi_0}\lambda^2A^2r^4+(3A^2e^{4\Phi_0}\lambda^2-e^{8\Phi_0}\lambda^4)r^2\\
&+e^{8\Phi_0}\lambda^4r b(r)-A^2e^{6\Phi_0}\lambda^6=0.
\end{aligned}
\end{equation}
This equation strictly depends on the functional form of the $b(r)$ function. The photon impact parameter $\lambda$ cannot assume any value, but it ranges in a limited interval. First of all, we must have $\lambda\ge0$ and since $\lambda$ depends both on the value of the source's radius $R_\star$ and angular velocity $\Omega_\star$, we should constraint such parameters. For reasons which will become clearer in Sec. \ref{sec:diagnostic}, we consider $R_\star=6M$, corresponding to the innermost stable circular orbit (ISCO) in the Schwarzschild metric. The  $\Omega_\star$ angular velocity has minimum and maximum values corresponding respectively to $\Omega_{\rm min}=0$ and $\Omega_{\rm max}=[\sqrt{-g_{tt}/g_{\varphi\varphi}}]_{r=R_\star}$. In the Schwarzschild metric we obtain $\Omega_\star\in[0,0.14] M^{-1}\ \rm{rad/s}$, or equivalently a rotation frequency of $f_\star\in[0,4126/(M/M_\odot)]$ Hz, which finally give $\lambda\in[0,7.35]\ M$.

For $\lambda/M\ll1$, considering only the terms less or equal than $\lambda^2$, from Eq. (\ref{eq:CH5}) we obtain
\begin{equation}
r^4-3e^{2\Phi_0}\lambda^2r^2+3e^{4\Phi_0}\lambda^2=0,
\end{equation}
where we have assumed that $e^{\Phi}$ and $b(r)$ are not functions of order higher than $\lambda^4$. Such equation admits no solutions, therefore we conclude that for $\lambda/M\ll1$, there are no critical hypersurfaces. The neglected terms would have given just a small contribution close to $r=0$, being still not an admissible solution for both a BH and a WH. As it will be clearer in Sec. \ref{sec:diagnostic}, we are only interested in solutions outside of the event horizon. 

Instead for $\lambda/M\gg1$ we have that Eq. (\ref{eq:CH5}) becomes
\begin{equation}
r^2-r b(r)+A^2e^{-2\Phi_0}\lambda^2=0,
\end{equation}
which depends on the explicit functional form of $b(r)$.

\section{Diagnostic to distinguish a black hole from a wormhole}
\label{sec:diagnostic}
In this section, we explain in details the strategy to diagnose a BH from a WH. The idea is mainly based on the hypothesis that a particular class of WH metrics admits a transition surface layer (located outside the event horizon), which is useful to smoothly connect the internal WH solution with the external WH region described by the Schwarzschild metric, see Sec. \ref{sec:geoastro} and Fig. \ref{fig:Fig2}. Therefore, in the WH case metric-changes occur in such transition surface layer, while in the BH case the metric continues to be that of the Schwarzschild spacetime.

We consider the presence of an accretion disk around the central compact object. It represents the source both of the radiation emission and of an intense magnetic field which produces squeezed vacuum states generating the negative energy required to make the WH both traversable and stable (see Sec. \ref{sec:exomat}). The general relativistic PR effect is very important for explaining the presence of stable critical hypersurfaces in the transition surface layer. We consider the emission proprieties not only from the disk (as it is usually done in the literature), but also from the critical hypersurface in the BH Schwarzschild metric (see Sec. \ref{sec:raytrace}). If a WH is present, we can have either no critical hypersurfaces or even if they exist, they have different emission proprieties from that of the Schwarzschild case, due to the presence of a different metric. Therefore, if we are able to fit the observational data through this model, it means that there is a BH, otherwise a WH could exist. \emph{We note that the flux emitted from the critical hypersurface in the Schwarzschild metric is a critical observable, which allows us to strongly reveal the presence of a BH.}

\subsection{Geometrical and astrophysical setup}
\label{sec:geoastro}
We consider a particular class of static and spherically symmetric WHs\footnote{We refer to the third example in the appendix of Ref. \cite{Morris1988}, entitled \emph{\qm{Solutions with exotic matter limited to the throat vicinity}}.}, contained in the appendix of the work by Morris and Thorne \cite{Morris1988}, which constitute subtle examples of perfect BH mimickers, indistinguishable both from X-ray electromagnetic and GW emissions. 

\begin{figure}[ht!]
\centering
\includegraphics[scale=0.52]{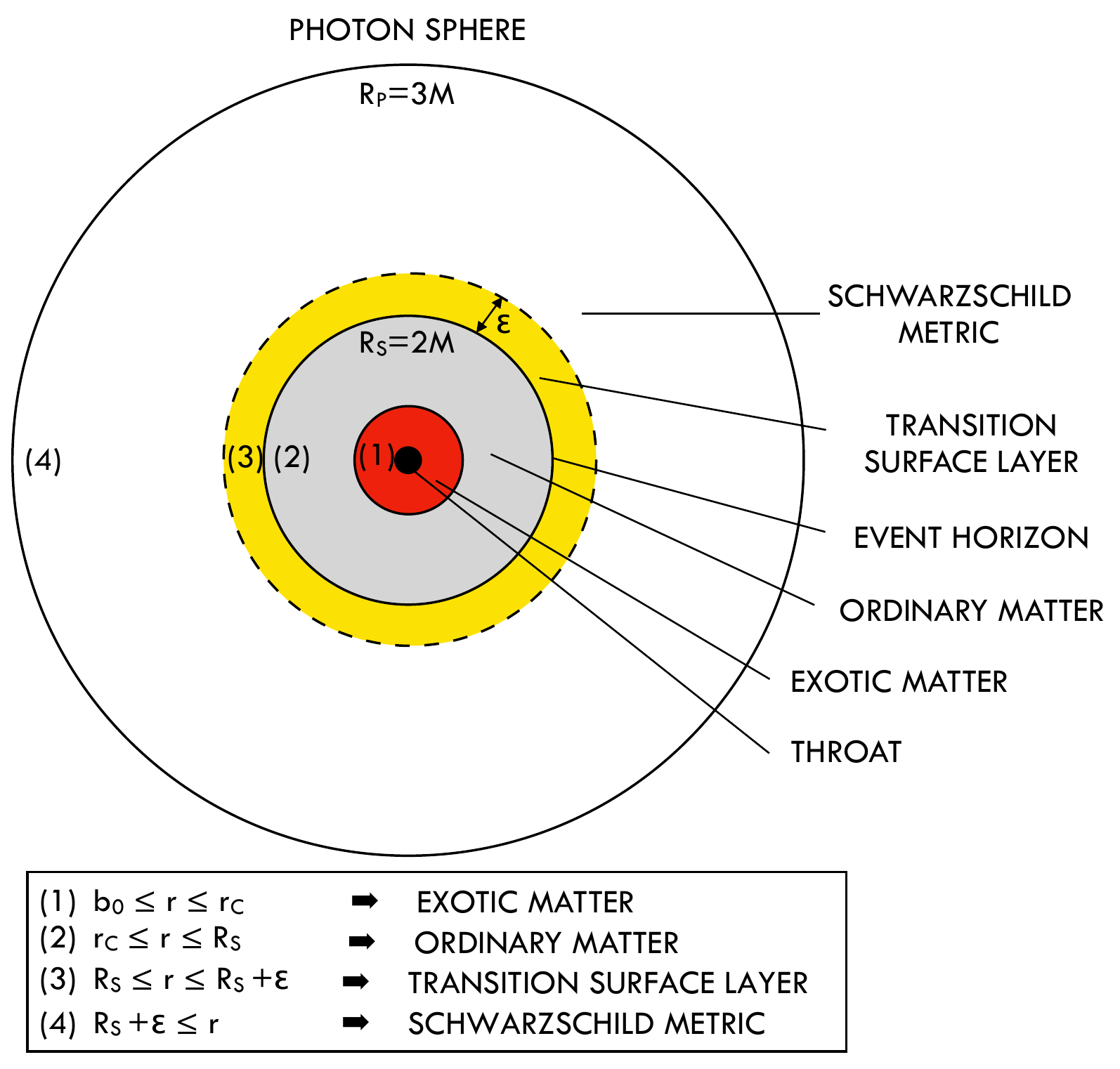}
\caption{Schematic representation of one of the Morris-Thorne solutions in terms of the spacetime domain.}
\label{fig:Fig2}
\end{figure}

In such WH metrics, the exotic matter is confined only in a small region close to the WH throat (i.e., $b_0\le r\le r_{\rm c}$), and outside there is ordinary matter extending up to the Schwarzschild radius $R_{\rm S}=2M$ (i.e., $r_{\rm c}\le r\le R_{\rm S}$). There exists a small transition surface layer (i.e., $R_{\rm S}\le r\le R_{\rm S}+\epsilon$) where it is possible to smoothly connect the inner WH solution with the Schwarzschild metric (i.e., $R_{\rm S}+\epsilon\le r$). Following the results reported by Cardoso and collaborators \cite{Cardoso2016}, we know that a good BH mimicker should have $R_{\rm S}+\epsilon<R_{\rm P}=3M$, corresponding to the photon sphere radius in the Schwarzschild metric. Based on this consideration, the transition surface layer must be located under the photon sphere radius, which consequently translates in having $\epsilon/M<1$ (see Fig. \ref{fig:Fig2}, for details). 

We note that the transition surface layer must be located outside the Schwarzschild event horizon, otherwise it would create a metric with a horizon, going therefore against the definition of WH. We have built up a very extreme case, which makes very complicate and thorny the identification of the presence of a WH, being in agreement with the actual state of the art of the observations. In addition, the location of the transition surface layer, which is very close to the event horizon, entails to inquire gravity in extreme field regimes.

In this geometrical background, we consider a thin accretion disk \cite{Shakura1973} located in the equatorial plane around the WH, extending from the Schwarzschild ISCO radius $r_{\rm in}=6M$ until $r_{\rm out}=100M$\footnote{The outer boundary of the disk mainly depends on the particular systems under investigation. We consider a sufficiently high value, because the main contributions derive from the inner regions closer to the BH, while the distant regions give only minor contributions, which do not drastically change our analysis.}. 
We assume that the accretion disk is present only in one universe, and not also in the opposite side. We exclude the possibility of two accretion disks in both universes, otherwise the WH can be immediately detected from X-ray observations \cite{Paul2019}. The accretion disk elements, which move down to the ISCO radius can be modeled as test particles, having as initial position $(r,\varphi)=(r_0,\varphi_0)$ and velocity $(\nu,\alpha)=(\nu_0,\alpha_0)$. They are influenced by the radiation field coming from the accretion disk, which in the 2D case can be reasonably approximated as a ring at ISCO radius, because the radiation field from other parts of the disk is shielded.
However, it is possible also to have other emitting sources, for example like a hot corona around a BH \cite{Fabian2015}.

\subsection{Magnetic field in the accretion disk as possible mechanism to make a BH traversable and stable}
\label{sec:exomat}
The presence of an accretion disk around a BH is an important source of information about the system under study. Indeed, the role of the accretion disk might have a twofold advantage: (1) it is the emitting source which generates a radiation field (source of the general relativistic PR effect), (2) its intense magnetic field makes a BH traversable and stable. About the latter issue, we propose a possible explanation of the WH stress-energy tensor (\ref{eq:setm}), and suggest a possible mechanism through which we can select among the plethora of known astrophysical BH systems the possible candidates hosting WHs. We do not enter into the modeling details of such process, because this goes beyond the aim of this paper.

Following an idea contained in the paper of Morris and Thorne \cite{Morris1988}, besides the possibility of modified gravity discussed above, we think that a situation where quantum fields can have negative energy density, violating thus the null energy condition (NEC), is obtained by a \emph{squeezed quantum state of the electromagnetic field}. Such phenomenon consists in decreasing the noise in one observable (coincident in our case with the energy) to consequently enhance the noise in the conjugate observable. The result is that the variations in the first observable are reduced below the quantum vacuum zero-point fluctuations, entailing thus negative energy \cite{Morris1988,Drummond2004,Davis2006}. In other words, quantum squeezing is useful to withdraw energy from one region standing in the ordinary vacuum at the cost of piling up the remaining energy elsewhere. In addition, such state is physically reproducible in laboratory thanks to the nonlinear-optics squeezing technique (see \cite{Davis2006}, and references therein), being, therefore, one of the most feasible astrophysical candidates.

We underline again that we are interested in the situation where an element of accreting gas can fall inside the WH, rather than a human traverses it for interstellar travels. We deem that the possible cause for generating the squeezed vacuum states is due to the presence of strong magnetic fields in the accretion disk \cite{Morris1988}. In addition, since such magnetic fields have steady strength, a WH both \emph{traversable and stable} could be realized. This is not just an hypothesis, because in the reality there exists astrophysical BH systems endowed with accretion disk structures showing intense magnetic fields \cite{Piotrovich2015}, such as BHs in active galactic nuclei (AGNs): NGC 7469 ($2.20\times10^5$ G), Akn 564 ($1.26\times10^5$ G), NGC 4051($9.85\times10^4$ G), PG 1211+143 ($6.25\times10^4$ G), Mrk 335 ($6.10\times10^5$ G). Adopting Eq. (2.11) in Ref. \cite{Shakura1973}, where $B\sim10^8(M_\odot/M)^{1/2}$ G at the ISCO radius, it is possible to obtain higher magnetic fields, as $B\sim10^7$ for $M=100M_\odot$, and $B\sim10^6$ for $M=10^4M_\odot$.

\subsection{Ray-tracing of emitting surfaces}
\label{sec:raytrace}
From Sec. \ref{sec:CH} we know that it is always possibile to have a critical hypersurface very close to the event horizon in the Schwarzschild metric, see Fig. \ref{fig:Fig4}. We study the emission proprieties from such configuration together with those of the accretion disk in the BH case toward a distant observer.  Such research topic has been only partially treated in the PR effect literature \cite{Bini2012}. 

The calculations of the emitted fluxes can be performed by exploiting the \emph{ray-tracing technique}, which relies on tracking a photon trajectory from the emission point to the observer location. In order to carry out such calculations in Schwarzschild metric, there are some fundamental effects to be taken into account, which are \cite{Misner1973,Defalco2016}: light bending, gravitational lensing (or better known as solid angle), and gravitational redshift. 
\begin{figure}[ht!]
\centering
\includegraphics[scale=0.56]{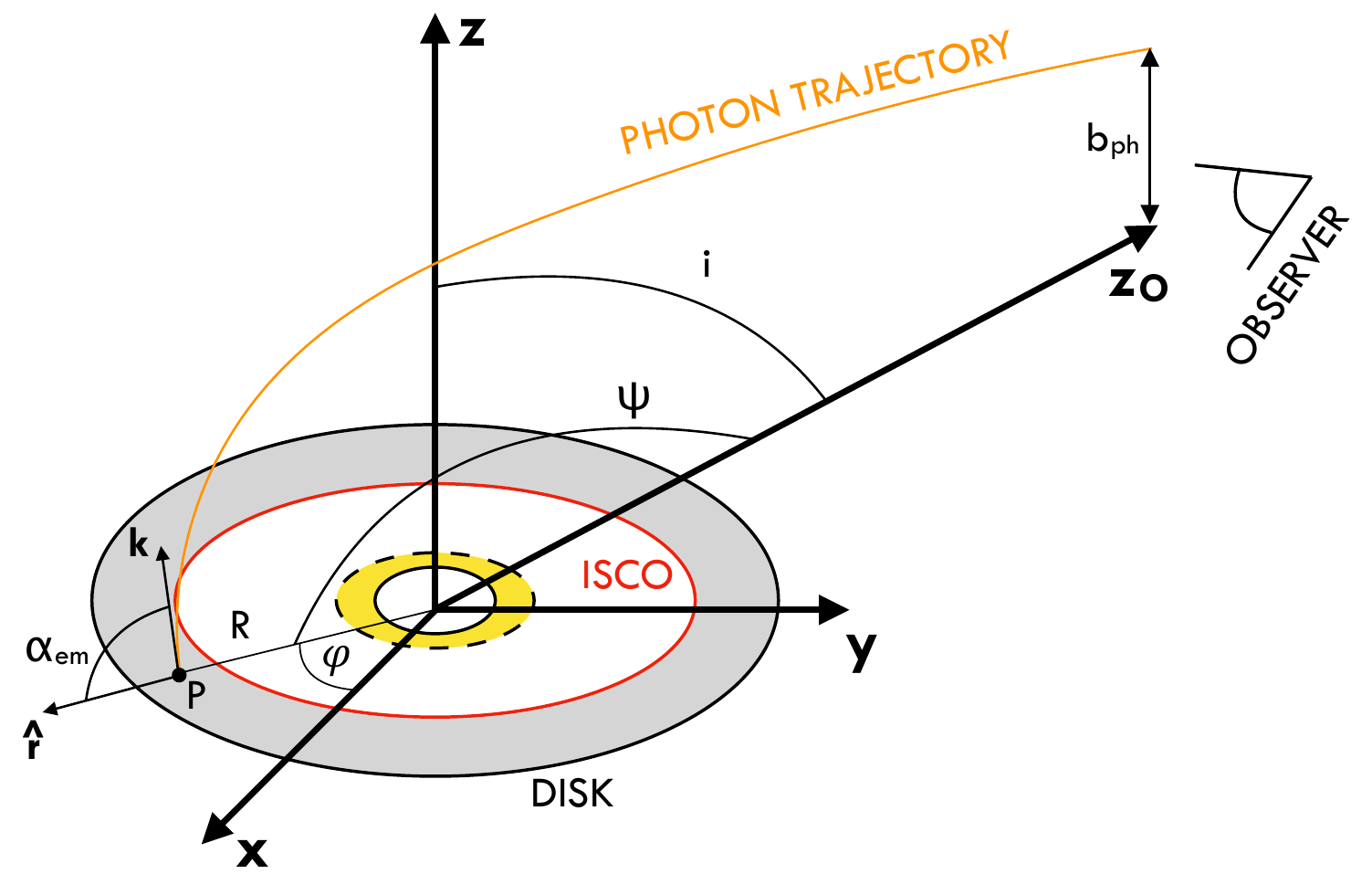}
\caption{Ray-tracing geometry in Schwarzschild spacetime.}
\label{fig:Fig3}
\end{figure}

The ray-tracing geometry is depicted in Fig. \ref{fig:Fig3}. We consider a reference frame centered at the origin of the BH/WH location and having the $\boldsymbol{x}$- and $\boldsymbol{y}$-axes lying in the equatorial plane, and the $\boldsymbol{z}$-axis orthogonal to the equatorial plane. We can adopt spherical coordinates, where the radius $r$ joins the center of the coordinates with any point in the space, the azimuthal angle $\varphi$ measured clockwise from the $\boldsymbol{x}$-axis, and the latitudinal angle $\theta$ measured from the $\boldsymbol{z}$-axis. There is a static and not rotating observer located at infinity, who is inclined by an angle $i$ with respect to the $\boldsymbol{z}$-axis, and the $\boldsymbol{z_0}$-axis points in the observer's direction. The latitudinal angle $\psi$ is measured from the $\boldsymbol{z_0}$-axis, and it is known in the literature as \emph{light bending angle} \cite{Misner1973,Beloborodov2002,Defalco2016}.

Let us consider a point $P$ in the space at radial distance $R$, where a photon is emitted (see Fig. \ref{fig:Fig3}). In such point, we can define the emission angle $\alpha_{\rm em}$ as the angle formed by the radial versor $\boldsymbol{\hat r}$ and the photon velocity $\boldsymbol{k}$ both applied in the point $P$. The photon will follow a null-trajectory in the Schwarzschild spacetime (lying in a single plane), which will reach the observer location with a photon impact parameter \cite{Misner1973}
\begin{equation} \label{eq:bpho}
b_{\rm ph}=\frac{R\sin\alpha_{\rm em}}{\sqrt{1-2M/R}}.
\end{equation}
It is important to note that this photon impact parameter $b_{\rm ph}$ is different from that of the radiation field $\lambda$, see Eq. (\ref{MT_impact_parameter}). 
We will employ high-accurate approximate polynomial ray-tracing equations for the accretion disk, while for the critical hypersurface we are forced to use original integral formulas since in that region are not available accurate approximate equations. 

We will produce some emission templates from the critical hypersurface (see Sec. \ref{sec:ECH}), accretion disk (see Sec. \ref{sec:EAC}), and combined profiles (see Sec. \ref{sec:ET}). Then, we will analyse their behaviors to extract relevant physical information. We will focus only on the Schwarzschild spacetime, because for other metrics in the surface transition layer (i.e., $2M\le r\le 2M+\epsilon$) it can occur either that there are no critical hypersurfaces, so the matter flows down to the throat, or even if they exist, they have different emission proprieties with respect to the Schwarzschild metric, which the observer at infinity can immediately distinguish (see Sec. \ref{sec:CH}). 

We do not perform the same calculations in the WH case, for mainly two reasons: (1) we do not know \emph{a-priori} the most suitable metric to be used in the transition surface layer, but instead it can be determined \emph{a-posteriori} if the observational data shows strong departures from the BH model; (2) the mathematical problem behind this case is very complex and can be the topic of another paper. Indeed, it entails to solve these issues: $(i)$ developing the ray-tracing equations in such new metric, $(ii)$ analysing their proprieties, $(iii)$ smoothly matching such equations with those of the Schwarzschild metric on the boundary of the transition surface layer.

\subsubsection{Emission from the critical hypersurface}
\label{sec:ECH}
\begin{figure}[ht!]
\centering
\includegraphics[scale=0.28]{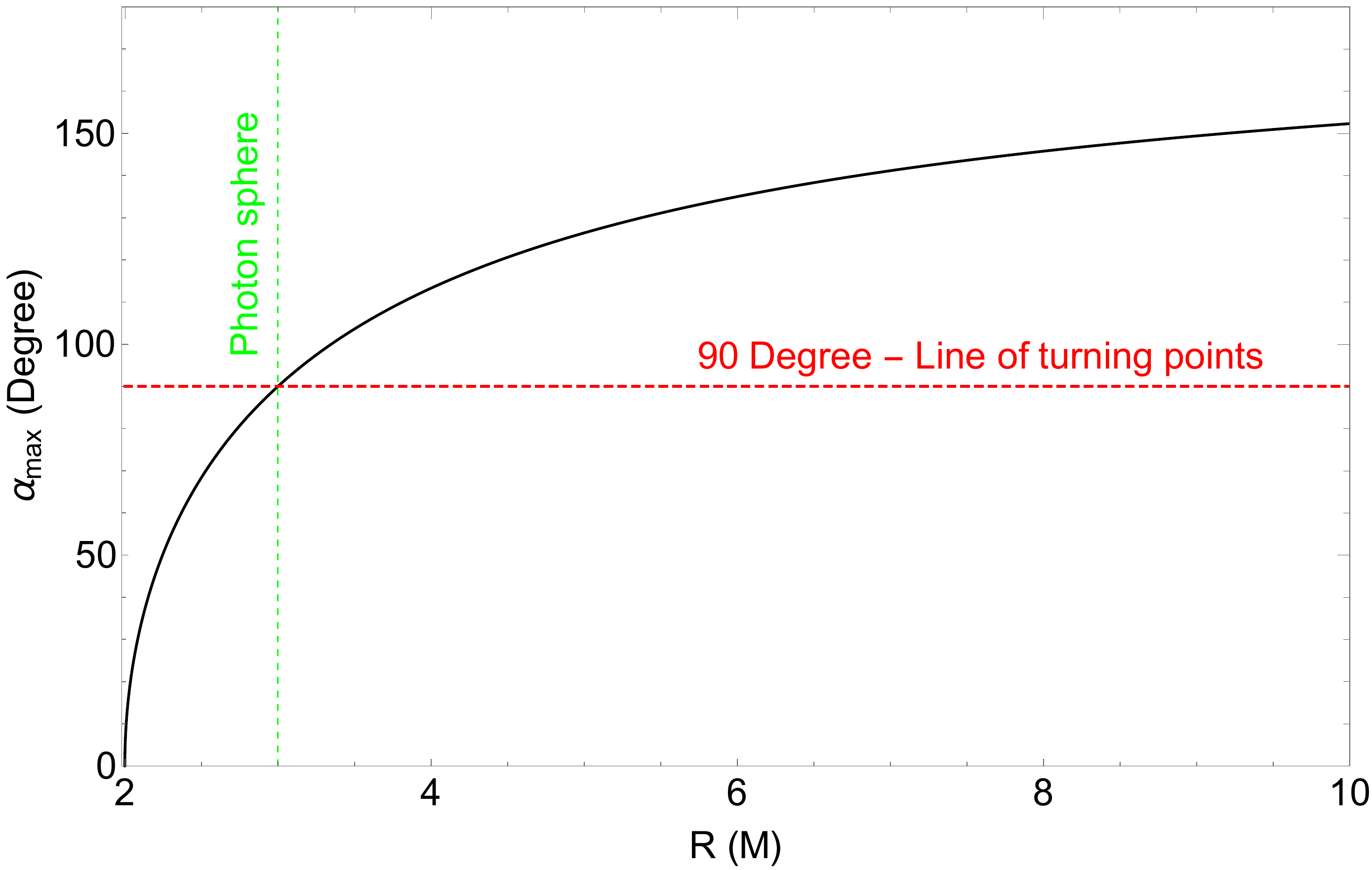}
\caption{Maximum emission angle $\alpha_{\rm max}$ in terms of the emission radius $R$. The dashed red line is for $\alpha=\pi/2$, and this is the threshold to have turning points, while the dashed green line is to determine the position of the photon sphere.}
\label{fig:Fig6}
\end{figure}

The ray-tracing of the critical hypersurface considers for each point the related light bending angle, which is calculated through the formula \cite{Defalco2016}
\begin{equation}
\cos\psi=\sin i\cos\varphi.
\end{equation}
The light bending equation \cite{Misner1973,Beloborodov2002,Defalco2016}
\begin{equation} \label{eq:libe} 
\psi=\int_R^\infty \frac{dr}{r^2}\left[\frac{1}{b_{\rm ph}^2}-\frac{1}{r^2}\left(1-\frac{2M}{r} \right) \right]^{-\frac{1}{2}},
\end{equation}
is valid for every $R>2M$ and $\alpha_{\rm em}\in[0,\pi/2]$. Through interpolation numerical methods we determine the emission angle $\alpha_{\rm em}$ \cite{Press2002}. In particular we must distinguish photons with zero and one turning points \cite{Defalco2016}. Defined $\psi_{\rm p}$ as the light bending angle corresponding to the emission angle $\alpha_{\rm em}=\pi/2$, we have that for $\psi\in[0,\psi_{\rm p}]$ there are zero turning points, while for $\psi\in[\psi_{\rm p},\psi_{\rm max}]$ we have one turning point, where $\psi_{\rm max}$ is the light bending angle corresponding to the maximum emission angle \cite{Defalco2016}
\begin{equation} \label{alphamax}
\alpha_{\rm max}=\pi-\arcsin\left[\frac{3}{2}\sqrt{3\left(1-\frac{2M}{R}\right)}\frac{2M}{R}\right].
\end{equation}   
Indeed, for $\alpha_{\rm em}\in[\alpha_{\rm max},\pi]$ the photon is swallowed by the BH and cannot reach the observer. Such argument is valid for $R\ge3M$ (disk case), while for $R<3M$ (our case), we have that $\alpha_{\rm max}$ becomes\footnote{For $R\le3M$, it occurs that $\alpha_{\rm max}\le\pi/2$ until to arrive to $\alpha_{\rm max}=0$ at $R=2M$, forming the so called \emph{cone of avoidance} \cite{Chandrasekhar1992}.} \cite{Chandrasekhar1992}
\begin{equation} \label{alphamax2}
\alpha_{\rm max}=\arcsin\left[\frac{3}{2}\sqrt{3\left(1-\frac{2M}{R}\right)}\frac{2M}{R}\right].
\end{equation}   
This remark is very useful not only to make smooth the $\alpha_{\rm max}$ function, as it can be seen in Fig. \ref{fig:Fig6}, but also to perform a correct ray-tracing procedure.  

For photons with one turning point, we apply a \emph{symmetrization process} by defining $\psi_{\rm S}=2\psi_{\rm p}-\psi$,  because Eq. (\ref{eq:libe}) is defined only for $\psi\in[0,\psi_{\rm p}]$. Then, we will obtain the emission angle $\alpha_{\rm S}\in[0,\pi/2]$, therefore to obtain the right emission angle $\alpha_{\rm em}$ corresponding to $\psi$, we use another \emph{symmetrization process}, i.e., $\alpha_{\rm em}=\pi-\alpha_{\rm S}$ \cite{Defalco2016}. In our case we do not consider any symmetrization process since $\alpha_{\rm em}\le\pi/2$. 

\begin{figure}[ht!]
\centering
\includegraphics[trim=0cm 2cm 0cm 1cm,scale=0.3]{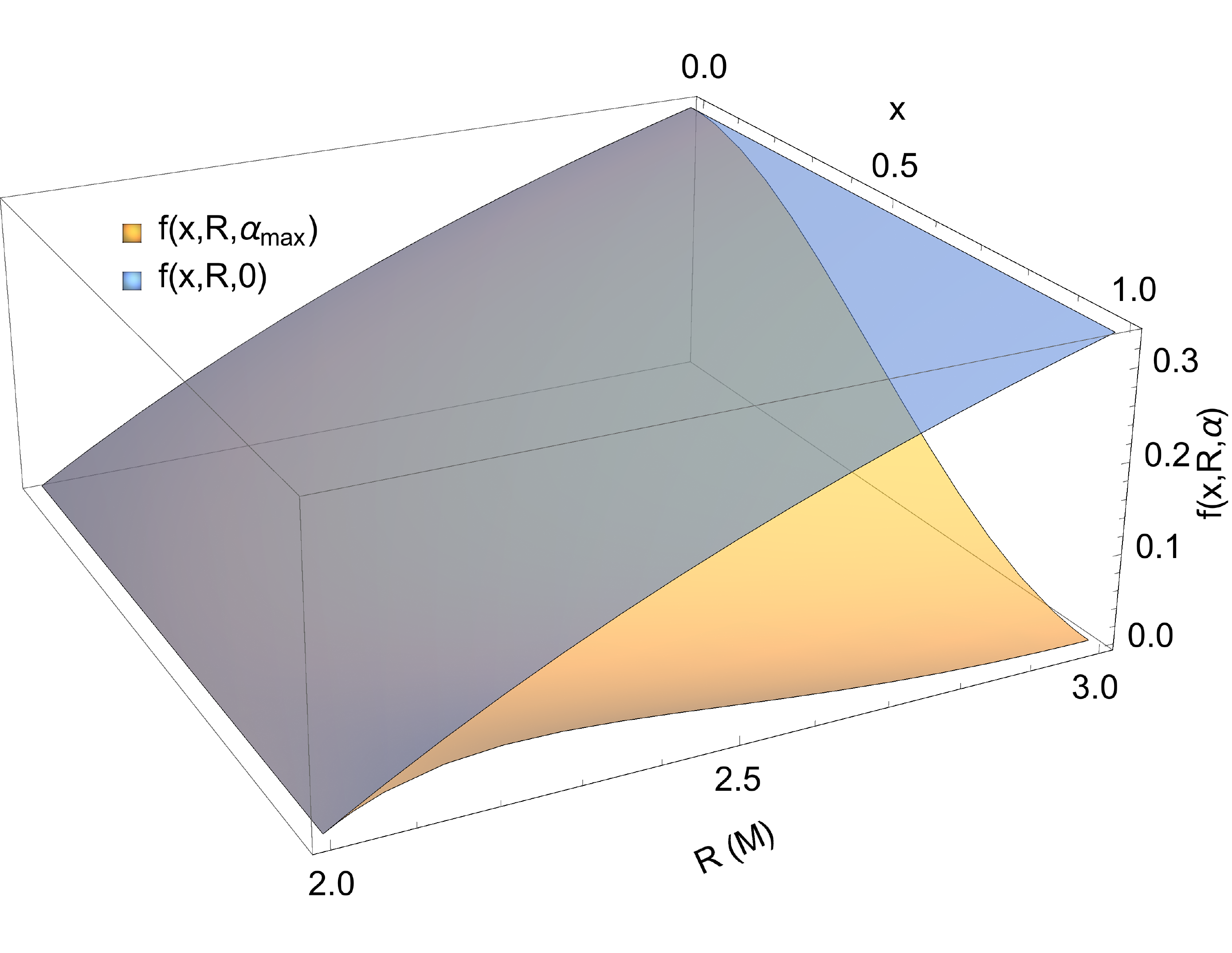}
\caption{Function $f(x,R,\alpha)$, see Eq. (53), plotted in terms of the emission radius $R$, and $x$ variable of integration. The blue and yellow surfaces are respectively for $f(x,R,0)$ and $f(x,R,\alpha_{\rm max})$.}
\label{fig:Fig7}
\end{figure}

Another non-trivial aspect is that in the range $R<3M$ the function in the square root of Eq. (\ref{eq:libe}) is always positive. For practical reasons, considering the change of variables $x=R/r$, we rewrite such equation as
\begin{equation} \label{eq:libe2} 
\psi=\int_0^1 \frac{\sin\alpha\ dx}{\sqrt{f(x,R,\alpha)}}.
\end{equation}
where 
\begin{equation} \label{eq:libe2} 
f(x,R,\alpha)=1-\frac{2M}{R}-x^2\sin^2\alpha\left(1-\frac{2Mx}{R} \right).
\end{equation}
In Fig. \ref{fig:Fig7} we prove what we have claimed.

We have all the elements for calculating the photon impact parameter $b_{\rm ph}$ and the solid angle formula \cite{Defalco2016}
\begin{equation} \label{EFSA} 
d\Omega=\frac{\frac{\cos i}{R^2\ \sin^2\psi}\frac{b_{\rm ph}^2}{\cos\alpha_{\rm em}}} {\int_R^\infty \frac{dr}{r^2}\left[1-\frac{b_{\rm ph}^2}{r^2}\left(1-\frac{2M}{r} \right) \right]^{-\frac{3}{2}}}\ dR\ d\varphi.
\end{equation}

For producing the emission profiles, we need only to calculate the gravitational redshift $(1+z)^{-1}$. The test particle velocity on the critical hypersurface with respect to the coordinate time $t$ is \cite{Bini2009,Bini2011}
\begin{equation}
U^\alpha\equiv\frac{dx^\alpha}{dt}=\left(1,0,0,\frac{R-2M}{R^3}\lambda\right).
\end{equation}
The photon velocity is \cite{Misner1973}
\begin{eqnarray}
k_t&=&-E,\\
k_r&=&E\sqrt{1-\frac{b_{\rm ph}^2}{R^2}\left(1-\frac{2M}{R}\right)}\left(1-\frac{2M}{R}\right)^{-1},\\
k_\varphi&=&Eb_{\rm ph},
\end{eqnarray}
 and the observer velocity is $V_0^\alpha=(1,0,0,0)$. Therefore, the gravitational redshift is \cite{Misner1973}
\begin{equation} \label{eq:redshift}
(1+z)^{-1}\equiv\frac{V_0^\alpha k_\alpha}{U^\alpha k_\alpha}=\left(1-\frac{R-2M}{R^3}\lambda b_{\rm ph}\right)^{-1}
\end{equation}

The flux emitted by the critical hypersurface for an observed frequency $\nu_{\rm em}$ can be calculated through \cite{Defalco2016}
\begin{equation} \label{eq:flux}
F_{\nu_{em}}=\int_{\Omega} \frac{\epsilon_0 \xi^{q}}{4\pi}\,(1+z)^{-4}\ d\Omega,
\end{equation}
where $\epsilon_0$ is the surface emissivity varying as a power law of $\xi=r/M$ with index $q$. 

\begin{figure}[ht!]
\centering
\includegraphics[trim=1.3cm 2cm 0cm 2.5cm,scale=0.35]{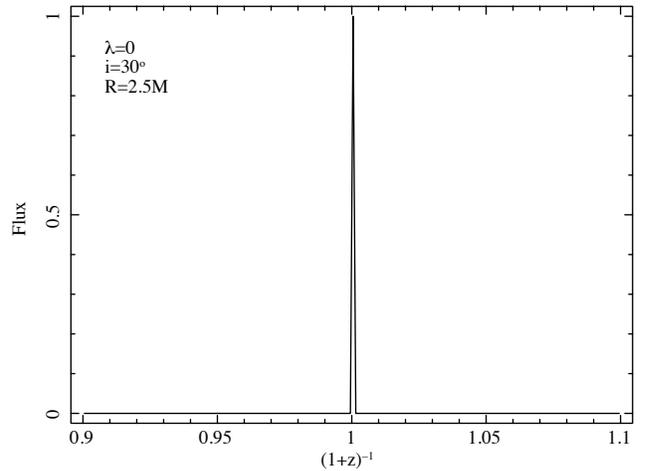}
\caption{Normalized flux of the critical hypersurface for $\lambda=0,i=30^\circ,R=2.5M$, and surface emissivity index $q =-3$.}
\label{fig:Fig8}
\end{figure}

For $\lambda=0$, we know that the test particle does not move, therefore for whatever emission radius $R$ and observer inclination angle $i$, we have a profile peaked at 1, as we expect and it can be seen in Fig. \ref{fig:Fig8}.

\begin{figure*}[p!]
\centering
\vbox{
\includegraphics[trim=1.3cm 2cm 0cm 0cm,scale=0.6]{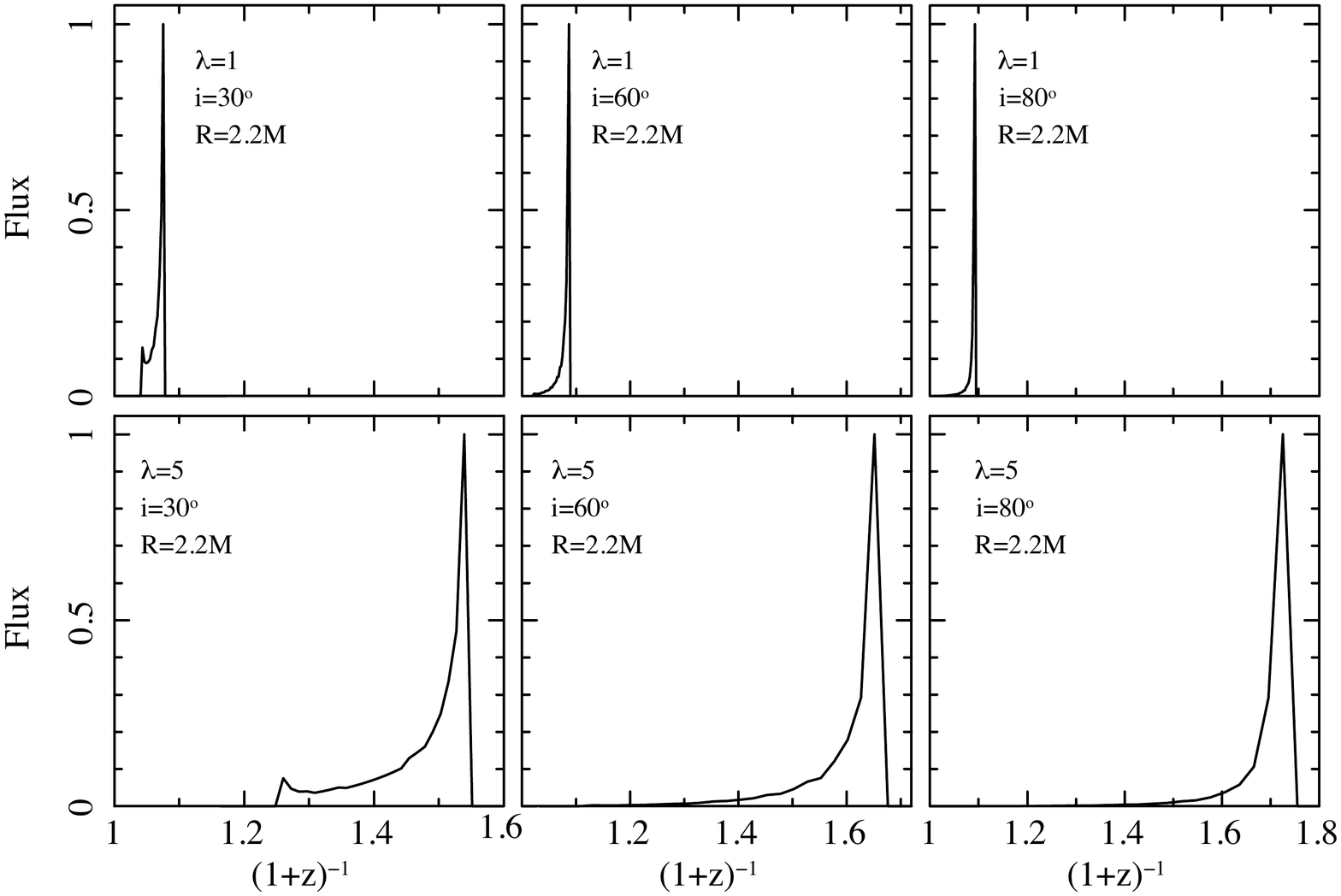}
\includegraphics[trim=1.3cm 2cm 0cm 1cm,scale=0.6]{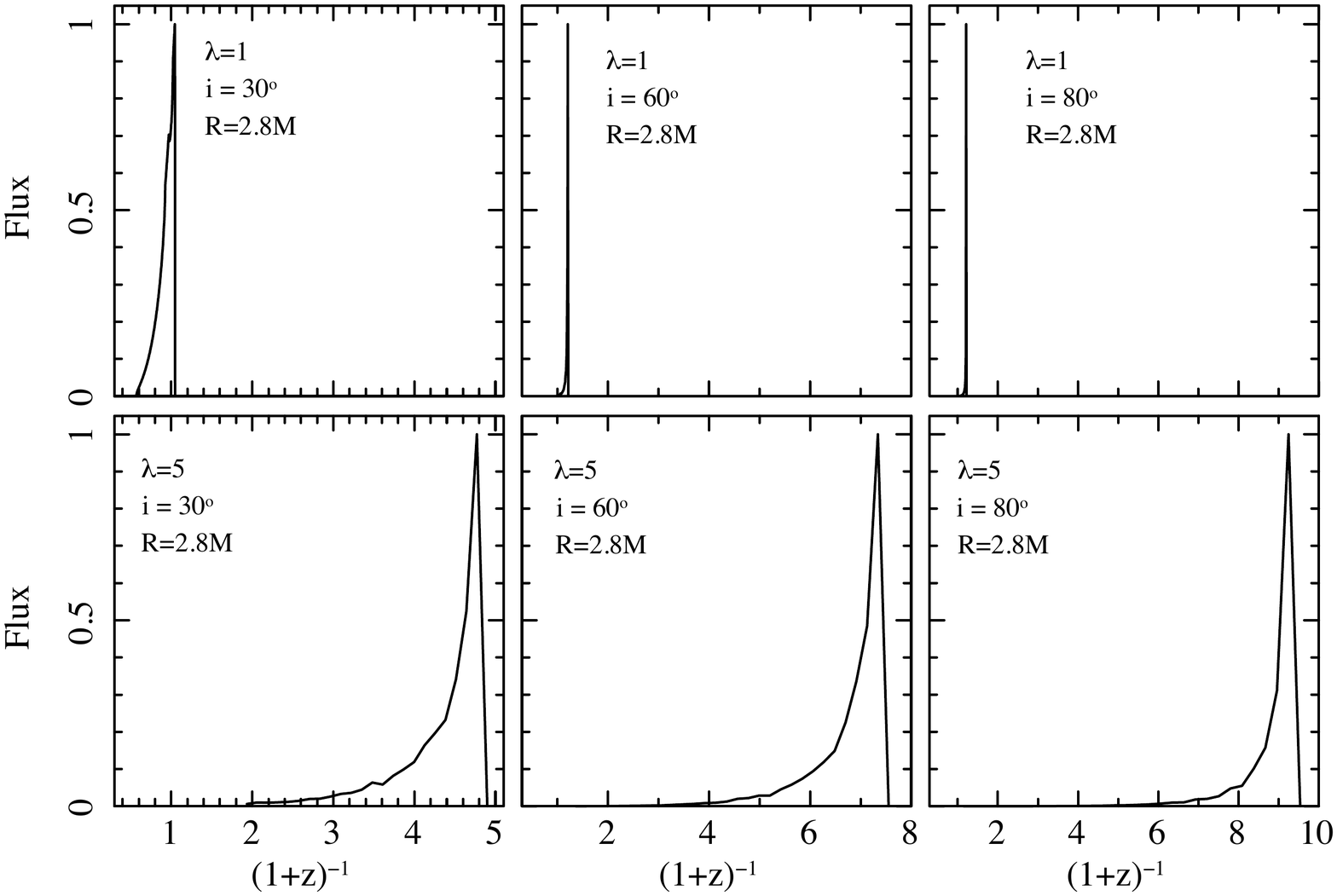}}
\caption{Critical hypersurface's normalized fluxes for $R=2.2M$, $\lambda=1,5$, $i=30^\circ,60^\circ,80^\circ$,and $q =-3$.}
\label{fig:Fig9}
\end{figure*}

In Fig. \ref{fig:Fig9} we display different templates performed for different $\lambda$ values (i.e., $\lambda=1,5$), observer inclination angles $i$ (i.e., $i=30^\circ,60^\circ,80^\circ$), and for emission radii ranging from very close to the event horizon ($R=2.2M$) to near the photon sphere ($R=2.8M$). 
These profiles are important to have information on the critical hypersurface's features in the BH case. They behave as broad iron line profiles shaped by the PR effect, where a fundamental parameter is the PR-radiation photon impact parameter $\lambda$. Indeed, for $\lambda/M\le1$ the fluxes peak very close to 1, while for $\lambda/M\ge1$ the fluxes depart from it and become broader. Since the gravitational redshift (\ref{eq:redshift}) depends only on the transverse velocity of the matter on the critical hypersurface, the higher is the velocity, the broader is the profile. The general relativistic effects are enhanced by increasing the observer inclination angle. For astrophysical purposes, it is useful to calculate for each emission profile (determined by $\lambda$ and $R=r_{\rm crit}$) the related luminosity parameter $A(r_{\rm crit},\lambda)/M$ through Eq. (\ref{eq:CH3}), see Table \ref{tab:Table1}. This information permits to have a list of luminosities emitted from astrophysical sources, which are very helpful both to obtain a set of input parameters for our model (i.e., $r_{\rm crit}$ and $\lambda$), and also viceversa to select the systems where to look for WHs.

\begin{table}[h] 
\caption{Different values of $A(r_{\rm crit},\lambda)/M$.}
\centering
\begin{spacing}{1.2}
\begin{tabular}{c||cccccccc} 
\hline 
\hline 
 &  &  &  &   &  &  & & \\
$r_{\rm crit}\setminus\lambda$ & 0 & 1 & 2 & 3 & 4 & 5 & 6 & 7\\
 &  &  &  &  &  &  & & \\
\hline 
$2.1M$ & 0.22 & 0.21 & 0.20 & 0.19 & 0.16 & 0.13 & 0.10 & 0.07 \\
$2.2M$ & 0.30 & 0.29 & 0.26 & 0.22 & 0.17 & 0.11 & 0.05 & 0.01 \\
$2.3M$ & 0.36 & 0.35 & 0.30 & 0.23 & 0.15 & 0.07 & 0.01 & --     \\
$2.4M$ & 0.41 & 0.39 & 0.32 & 0.23 & 0.13 & 0.04 & --        & --   \\
$2.5M$ & 0.45 & 0.42 & 0.34 & 0.23 & 0.11 & 0.02 & --        & --   \\
$2.6M$ & 0.48 & 0.45 & 0.35 & 0.23 & 0.10 & 0.01 & --        & --   \\
$2.7M$ & 0.51 & 0.47 & 0.36 & 0.22 & 0.09 & 0.01 & --        & --   \\
$2.8M$ & 0.54 & 0.49 & 0.37 & 0.22 & 0.08 & 0.00 & --        & --   \\
$2.9M$ & 0.56 & 0.51 & 0.38 & 0.21 & 0.07 & 0.00 & --        & --   \\
\hline  
\hline  
\end{tabular} 
\end{spacing}
\label{tab:Table1}
\end{table} 

\subsubsection{Emission from the disk}
\label{sec:EAC}
We consider the emission from the disk modeled by the very broad iron line (Fe K$_{\rm\alpha}$) observed around 6.4 keV in a number of astrophysical systems (see Refs. \cite{Miller2007,Cackett2010}, and references therein), well known in the high-energy literature \cite{Frank2002}. Since the accretion disk extends from $r\ge R_{\rm ISCO}$, it is sufficient to employ the approximate polynomial-formulas by De Falco and collaborators \cite{Defalco2016}\footnote{There are other more accurate formulas proposed in the literature, which can be also employed, although they are extremely useful when we are closer to the photon sphere location \cite{Semerak2015,Laplaca2019,Poutanen2019}. In our case, the accuracy of our approximation is enough \cite{Defalco2016}.}. 

We apply the same ray-tracing procedure scheme discussed in the previous section, but we replace the light bending equation (\ref{eq:libe}) with \cite{Beloborodov2002,Defalco2016}
\begin{equation}
\alpha_{\rm em}=\arccos\left[1-(1-\cos\psi)\left(1-u\right)\right].
\end{equation}
where $u=2M/r$. Such formula permits to easily determine $\alpha_{\rm em}$, without resorting to any numerical method. We replace also the solid angle formula (\ref{EFSA}) with \cite{Defalco2016}
\begin{equation} \label{AFSA} 
\begin{aligned}
d\Omega&\approx \frac{\cos i}{\sin^2\ \psi\ (1-u)}\ R \left[ 2z+\left(1-2C\right)z^2+\right. \\
&\left. +\left(1-C+2C^2-2D\right)z^3\right]\ dR\ d\varphi,
\end{aligned}
\end{equation}
where
\begin{equation}
C=\frac{4-3u}{4(1-u)},\qquad D=\frac{39u^2-91u+56}{56(1-u)^2}.
\end{equation}

The gravitational redshift (\ref{eq:redshift}) is also replaced with
\begin{equation} \label{eq:redshift2}
\begin{aligned}
(1+z)^{-1}=\frac{\left(1-\frac{2M}{r}-\omega_k^{2}r^{2}\right)^{1/2}}{\left(1+b_{\rm ph}\omega_k         
\frac{\sin i\,\sin\varphi}{\sin\psi} \right)}, 
\end{aligned}
\end{equation}
where we consider matter orbiting with Keplerian velocity $\omega_k=\sqrt{M/r^3}$ around the BH. The observed flux is still calculated by using Eq. (\ref{eq:flux}).
\begin{figure}[ht!]
\centering
\includegraphics[trim=1.3cm 2cm 0cm 0cm,scale=0.4]{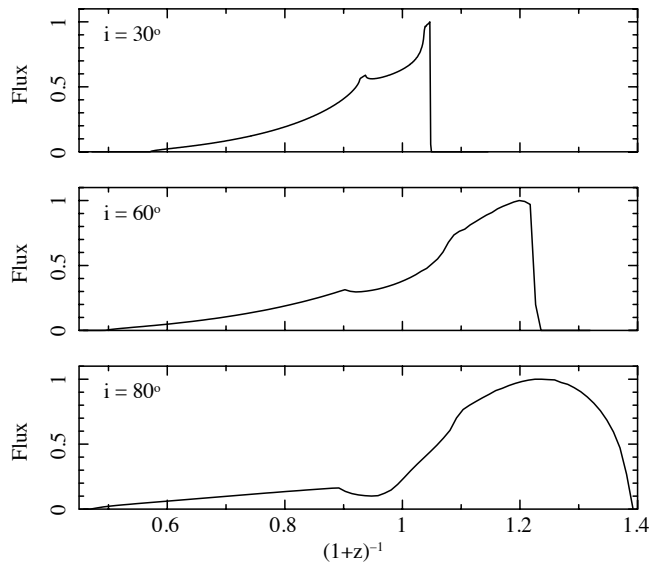}
\caption{Normalized broad iron line profiles for isotropic radiation emission from a disk extending from $r_{\rm in} =6M$ to $r_{\rm out} =100M$, assuming a surface emissivity index $q =-3$ and observer inclination angles $i=30^\circ,60^\circ,80^\circ$.}
\label{fig:Fig5}
\end{figure}

In Fig. \ref{fig:Fig5}, we show some disk emission plots, exhibiting the characteristic skewed, highly broadened, and double-horned line profiles \cite{Fabian1989,Bao1994,Beckwith2004}. The general relativistic effects, together with the transversal Doppler shifts and gravitational redshift, strongly shape the iron K$_{\alpha}$ line, allowing thus to inquire the accretion disk dynamics \cite{McClintock2011}. The general relativistic effects are enhanced by increasing the observer inclination angle $i$. The highest peak corresponds to the blue-shifted emission from material on the approaching side, while the other peak is related to red-shifted emission from matter on the receding part. The broadest part of the line is produced by the fastest motion of matter in the inner regions \cite{Fabian1989}. 

\subsubsection{Total emission}
\label{sec:ET}
Combining the results obtained for the critical hypersurface (Figs. \ref{fig:Fig8} and \ref{fig:Fig9}) with those of the disk (Fig. \ref{fig:Fig5}), we can produce the total emission from astrophysical systems hosting BHs, which is what we actually observe. 

\begin{figure}[ht!]
\centering
\includegraphics[trim=1.3cm 2cm 0cm 3.5cm,scale=0.35]{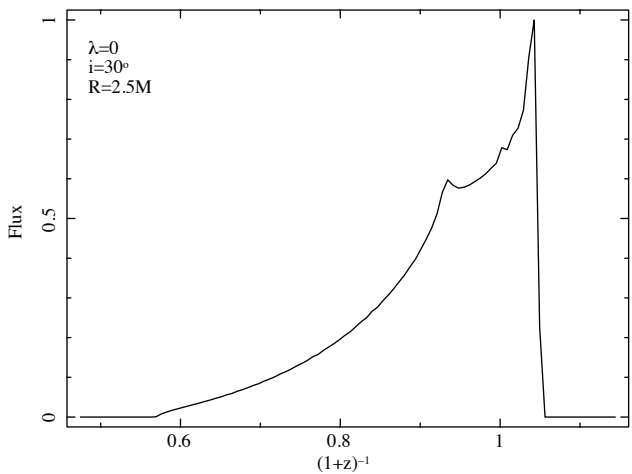}
\caption{Normalized flux of the total system, using the data of Fig. \ref{fig:Fig8} for the critical hypersurface and Fig. \ref{fig:Fig5} for the disk.}
\label{fig:Fig10}
\end{figure}

For $\lambda=0$, we notice from Fig. \ref{fig:Fig10} a small intermediate peak (related to the critical hypersurface) between the other two peaks of the accretion disk. The characteristic shape of this system will be a \emph{three-horned profile}, where a peak gives important information on the critical hypersurface, and therefore on the metric where it moves. In this particular case, the flux from the critical hypersurface sums with those from the accretion disk, explaining the existence of the intermediate peak. 

\begin{figure*}[p!]
\centering
\vbox{
\includegraphics[trim=1.3cm 2cm 0cm 0cm,scale=0.6]{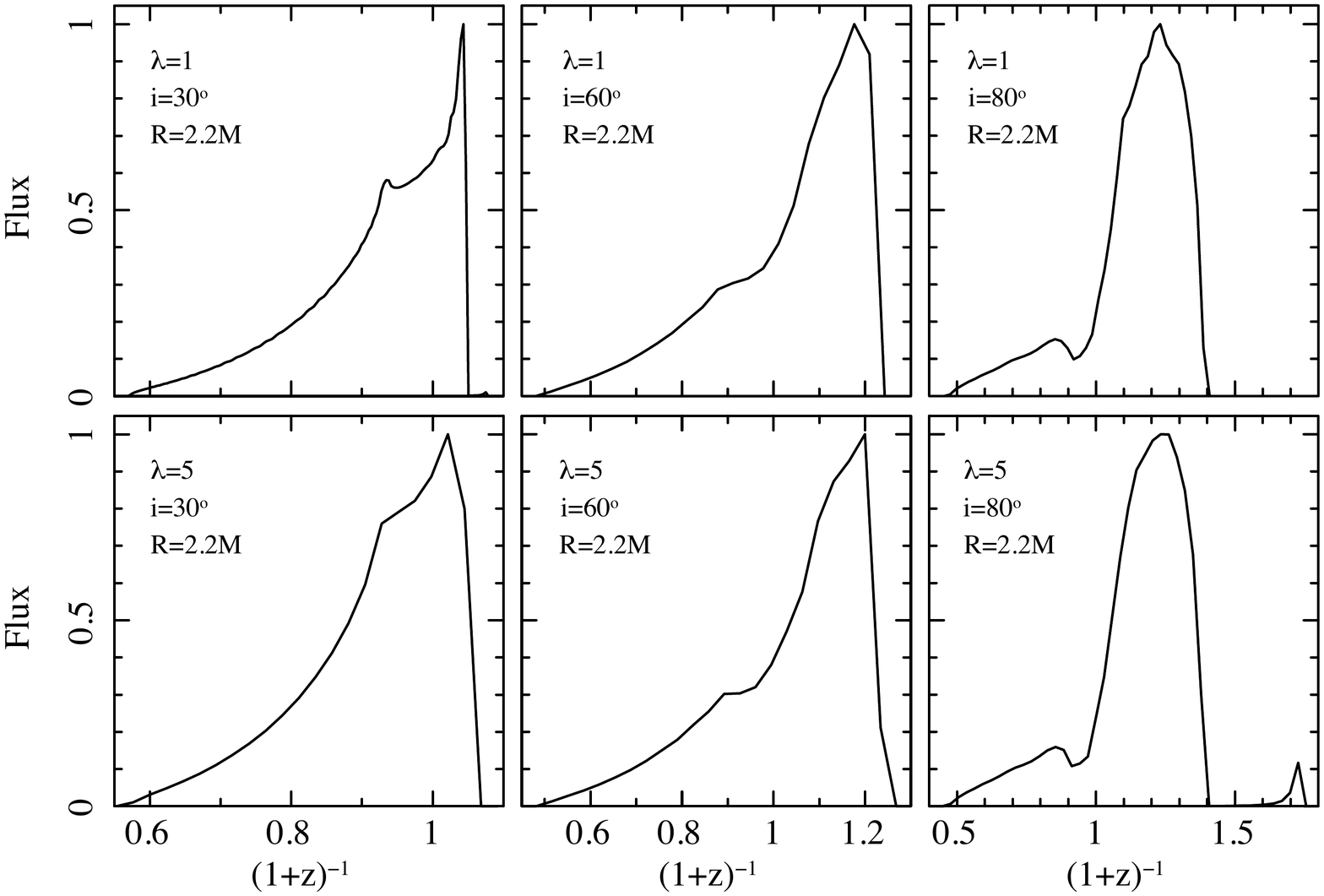}
\includegraphics[trim=1.3cm 2cm 0cm 1cm,scale=0.6]{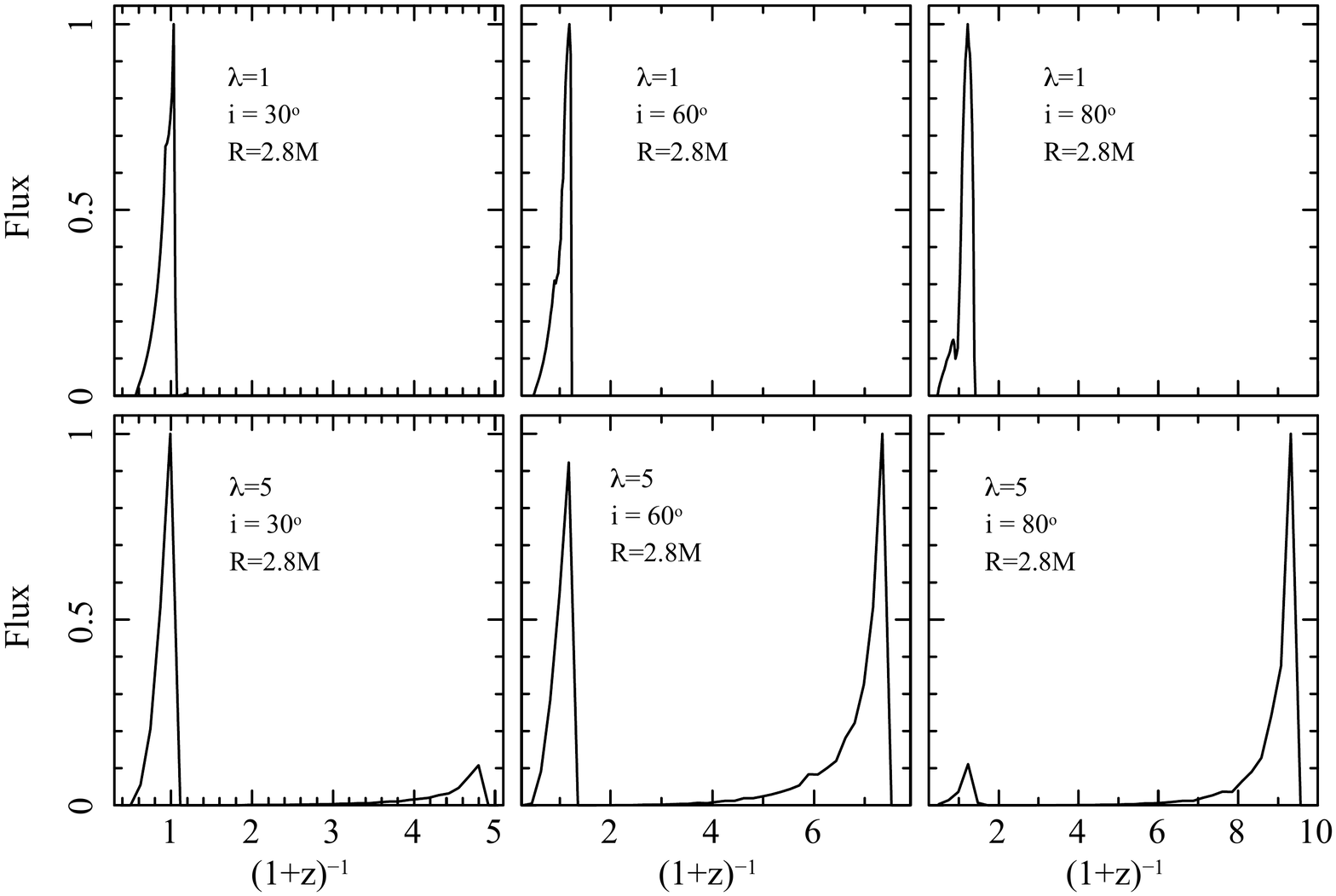}}
\caption{Normalized flux of the total system, using the data of Fig. \ref{fig:Fig9} for the critical hypersurface and Fig. \ref{fig:Fig5} for the disk.}
\label{fig:Fig11}
\end{figure*}

In Fig. \ref{fig:Fig11} we display a great variety of total emission profiles, combining the profiles from Figs. \ref{fig:Fig5} and \ref{fig:Fig9}. We have disparate behaviors, all showing in a way more or less pronounced the distinctive feature of the three-horned line.
Remembering that $\lambda$ is connected with the velocity of the matter moving on the critical hypersurface, we have:
for high values of $r_{\rm crit}$ and $\lambda$ the peak related to the PR effect is shifted to high energies and is more enhanced than those from the accretion disk, while for small values of both parameters the flux from the critical hypersurface sums with those of the disk. In this last cases, we have disk-like emission profiles with the presence of a distinctive higher blue-shifted peak.  

This new procedure can be heavily and timely supported by both present, like XMM-Newton \cite{Beckwith2004,Tomsick2014}, EHT \cite{Chael2016}, and future observational X-ray data, like Advanced Telescope for High-ENergy Astrophysics (ATHENA) \cite{Barcons2017}, enhanced X-ray Timing and Polarimetry mission (eXTP) \cite{Zhang2016}, Imaging X-ray Polarimetry Explorer (IXPE) \cite{Soffitta2013}. In addition, such technique could be simultaneously combined with other different methods of observations, like GWs detections or imaging of matter close to a BH, to acquire more and precise information. 

\section{Conclusions}
\label{sec:end}
In this work, we have derived the equations of motion (\ref{EoM1}) -- (\ref{time}) of a test particle in a static and spherically symmetric WH spacetime (Morris-Thorne like, see Sec. \ref{sec:MTmetric}) under the influence of the general relativistic PR effect, see Sec. \ref{sec:GRPReffect}. We consider a particular BH mimickers' class of WHs, where the exotic matter is placed in a small region close to the throat, and then ordinary matter extending up to $R_{\rm S}=2M$. A small transition surface layer located within $R_{\rm p}=3M$, permits to smoothly connect the inner solution to the Schwarzschild metric, see Fig. \ref{fig:Fig2}. A particular WH dynamics is determined, whenever the redshift $\Phi(r)$ and shape $b(r)$ functions are explicit. This dynamical system can admit the existence of a critical hypersurface, a region where there is an equilibrium between the radiation and gravitational forces, see Sec. \ref{sec:CH}. We have recalled some useful proprieties of such stable configurations in the Schwarzschild metric (see Fig. \ref{fig:Fig4}), and then we have investigated some general aspects for $\Phi=\Phi_0\equiv{\rm const}$ (value usually assumed by the redshift function in the transition surface layer \cite{Morris1988}). We have found that there are no critical hypersurfaces for $\lambda/M\ll1$, while for $\lambda/M\gg1$ it strongly depends on the functional form of the shape function $b(r)$. 

We have developed a diagnostic to distinguish a BH from the class of WHs outlined above. We have considered as astrophysical setup an accretion disk around a BH/WH (in only one universe) extending from the ISCO radius to $r_{\rm out}=100M$, see Sec. \ref{sec:geoastro}. The ISCO radius could be considered as the radiation source which alters the geodesic motion of the test particle through the general relativistic PR effect. We deem that the presence of an accretion disk with high magnetic fields $B\sim10^4-10^7$ G, might be the cause for producing squeezed vacuum electromagnetic states, phenomenon which generates negative energy and makes the BH traversable and stable, see Sec. \ref{sec:exomat}. This might be a very useful discriminant to reduce the search for WHs among several astrophysical systems. 

Knowing that the critical hypersurface can be located very close to the event horizon of a BH (or in the transition surface layer of a WH), this configuration can be employed to inquire the proprieties of the geometrical background and distinguish the two structures. An observable through which we can achieve such objective is the observed emission profiles, by employing the ray-tracing technique from the emission location toward the observer at infinity, see Sec. \ref{sec:raytrace}. We have first analysed the emission proprieties from the critical hypersurfaces in the BH Schwarzschild metric (see Figs. \ref{fig:Fig8} and \ref{fig:Fig9}), where they strongly depend on the $\lambda$ values, the critical hypersurface radius $r_{\rm crit}$, and the observer inclination angle $i$, which all contribute to enhance the general relativistic effects, see Sec. \ref{sec:ECH}. In such analysis, we have also provided the Table \ref{tab:Table1}, where we show some possible luminosities of BH systems, both for relating them with the input parameters of our model (i.e., $r_{\rm crit}$ and $\lambda$), and for reducing the astrophysical systems, where to look for WHs. Then, we have reported the emission from the accretion disk (see Fig. \ref{fig:Fig5}), exhibiting the characteristic skewed, double-horned iron line profile, see Sec. \ref{sec:EAC}. Finally, combining the two emissions to calculate the \emph{total} flux, we see the characteristic feature of a three-horned profile (see Figs. \ref{fig:Fig10} and \ref{fig:Fig11}), where a peak easily identifiable gives important information not only on the critical hypersurface, but also on the metric where the matter moves. Indeed, if the observational data can be well fitted by this model, we can conclude that there is a BH; instead if the fit through such description is not suitable to interpret the observational data, it means that there are metric-changes and the possibility to have a WH could be realistic, see Sec. \ref{sec:ET}. In addition, such method can be advantageously supported by the recent and near-future observational data, see Sec. \ref{sec:ET}. 

This work finds not only a practical implication in diagnosing a BH from a WH, but also other interesting applications. First of all, since the critical hypersurface lies very close to the event horizon, it can be extremely useful to infer fundamental proprieties on the gravitational field and also on how gravity couples with photons in strong field regimes. Another original idea consists in employing some generic spherically symmetric BH metrics, which are built within a model-independent framework and do not reflect nor need a specific theory of gravity \cite{Rezzolla2014}. They can be used to approximate arbitrary BH spacetimes through a small set of coefficients, that can be recovered from astronomical observations. In this way, we can measure in an agnostic manner possible deviations from GR and hence determine whether alternative theories of gravity are needed. In addition, this approach can be a tool to determine the WH metric in the transition surface layer, if one observes metric-changes there.

Finally, we stress that the method devised in this paper is based upon a \emph{toy model}, where several
elementary features can be further improved. Indeed as future projects, we aim at extending such description both in the diagnostic procedure and in the modeling aspects by improving the following aspects: considering a rotating axially symmetric WH spacetime in GR, eventually setting such description in the 3D space, and extending such treatment in modified theories of gravity. 

\section*{Acknowledgements}
V.D.F. thanks Osservatorio Astronomico Monte Porzio Catone for hospitality and support, and the Silesian University in Opava for support. V.D.F. and E.B. thank Gruppo Nazionale di Fisica Matematica of Istituto Nazionale di Alta Matematica for support. V.D.F. thanks Prof. Luigi Stella for the useful discussions. V.D.F. and E.B. thank Dr. Viacheslav Emelyanov for the useful suggestions and references on the squeezed vacuum states. S.C. and M.D.L. acknowledge the support of Istituto Italiano di Fisica Nucleare (INFN) {\it iniziative specifiche} MOONLIGHT2 and TEONGRAV.

\bibliography{references}
\end{document}